\documentclass[a4paper,11pt]{article}

\usepackage{jheppub}
\usepackage[T1]{fontenc}

\usepackage{graphicx} 
\usepackage{bm} 
\usepackage{epsfig}
\usepackage{amsmath, amssymb}
\usepackage{setspace}
\usepackage{color}
\usepackage{colordvi}
\usepackage{comment}
\usepackage{mystyle}

\newcommand{\beq}{\begin{equation}}
\newcommand{\eeq}{\end{equation}}
\newcommand{\bea}{\begin{eqnarray}}
\newcommand{\eea}{\end{eqnarray}}
\newcommand{\bit}{\begin{itemize}}
\newcommand{\eit}{\end{itemize}}

\newcommand{\nn}{\nonumber}

\def\calJ{{\cal J}}
\def\calo{{\cal O}}
\def\cals{{\cal S}}

\title{\boldmath Gravitational self-force in the ultra-relativistic limit: The ``large-$N$'' expansion}

\author[a]{Chad R.\;Galley,}
\author[b]{Rafael A.\;Porto}

\affiliation[a]{Theoretical Astrophysics, California Institute of Technology, Pasadena, CA 91125, USA}
\affiliation[b]{School of Natural Sciences, Institute for Advanced Study, Einstein Drive, Princeton, NJ 08540, USA}

\emailAdd{crgalley@tapir.caltech.edu}
\emailAdd{rporto@ias.edu}

\abstract{
We study the gravitational self-force using the effective field theory formalism. We show that in the ultra-relativistic limit $\gamma \to \infty$, with $\gamma$ the boost factor, many simplifications arise. Drawing parallels with the large $N$ limit in quantum field theory, we introduce the parameter $1/N \equiv 1/\gamma^2$ and show that the effective action admits a well defined expansion in powers of $\lambda \equiv N \epsilon$ at each order in $1/N$, where $\epsilon \equiv E_m/M$ and $E_m=\gamma m$ is the (kinetic) energy of the small mass. Moreover, we show that diagrams with nonlinear bulk interactions first enter at $\calo(\lambda^2/N^2)$ and only diagrams with nonlinearities in the worldline couplings, which are significantly easier to compute, survive in the large $N$/ultra-relativistic limit. Finally, we derive the self-force to $\calo (\lambda^4/N)$ and provide expressions for some conservative quantities for circular orbits. 
}

\begin{document}
\maketitle
\flushbottom

\newpage
\section{Introduction}
\label{sec:intro}

During the last several years a new formalism has emerged, based on effective field theory (EFT) ideas borrowed from particle physics, to study binary systems in general relativity. Originally the EFT approach was introduced within the post-Newtonian approximation for non-spinning \cite{Goldberger:2004jt} and spinning \cite{Porto:2005ac} inspirals, and has since produced a number of results for gravitationally interacting extended objects \cite{Goldberger:2004jt,Porto:2005ac, Gilmore:2008gq, Foffa:2011ub, Goldberger:2009qd, Porto:2010tr, Foffa:2011np, Galley:2012qs, Porto:2006bt, Porto:2008tb, Porto:2008jj, Porto:2012as, Goldberger:2005cd, Porto:2007qi, Porto:2010zg,Levi:2011eq, Kol:2011vg, Foffa:2012rn,Goldberger:2012kf}. Meanwhile, EFT ideas were also applied (besides in particle physics) to different areas, such as cosmology \cite{Cheung:2007st, Baumann:2010tm,LopezNacir:2011kk}, electrodynamics \cite{Galley:2010es}, fluid dynamics \cite{Endlich:2010hf,Endlich:2012vt, Grozdanov:2013dba}, and in particular to extreme mass ratio inspirals (EMRIs) \cite{Galley:2008ih, Galley:2010xn, Galley:2011te}, which is the subject of this paper.

The study of the self-force problem within the EFT approach was initiated in \cite{Galley:2008ih} where power-counting and leading order effects were worked out and a proof of the effacement of internal structure for EMRIs was given. Binaries with small mass ratios are often studied using perturbation theory performed in powers of $q \equiv m/M$ where $m$ represents a small mass object orbiting a much larger black hole with mass $M$. More generally, the expansion parameter is the size of the small object $R$ divided by the curvature length scale of the background spacetime. For EMRIs, these are $\sim m$ and $\sim M$, respectively. To date, only second-order $\calo(q^2)$  equations of motion are known at a formal level \cite{Gralla:2012db, Pound:2012nt}. 

In this paper we study the ultra-relativistic limit of the self-force problem where the boost factor $\gamma$ goes to infinity. An ultra-relativistic regime can be reached in several cases, such as circular orbits approaching the light ring in a black hole spacetime (the fact that these orbits are unstable is largely irrelevant for our theoretical study here), fast ``fly-by'' trajectories, and more generally fast moving objects in curved backgrounds. Inspired by an analogy with the large $N$ limit in quantum field theory \cite{tonyN}, we show here that many simplifications arise in the ultra-relativistic limit that are not captured by the canonical $m/M$ power-counting. We show that, upon introducing the expansion parameter $1/N \equiv \gamma^{-2}$ and defining $\lambda \equiv N \epsilon$ with $\epsilon \equiv E_m/M$ and $E_m=\gamma m$, the gravitational effective action (which yields the self-force) admits an expansion of the type
\beq
S_{\rm eff} = L/N \left( 1+ \lambda + \lambda^2+\ldots\right) + \calo(\lambda^2/N^2),
\eeq
where $L \sim E_m M (=\gamma m M)$ is the angular momentum of the small mass (in $G_N=c=1$ units used throughout). A similar expansion applies to the one-point function, $h_{\mu\nu} (x)$, which can also be used to compute the self-force as we discuss in this paper. Our goals here are: 1) to derive the new power-counting rules in the large $N$ limit; 2) to show that diagrams with nonlinear bulk interactions are subleading in the $1/N$ expansion; 3) to report the gravitational self-force to \emph{fourth} order in $\lambda$ at leading order in $1/N$; and 4) to provide formal expressions for conservative quantities for the particular case of circular orbits.
We conclude on a more formal note with some comments on the problem of finding the self-force in the exact massless limit, e.g. a photon moving in a black hole spacetime. 

\section{Power counting rules}
\label{sec:powercounting}

Our setup is the same as in the standard EMRI EFT \cite{Galley:2008ih} except that we consider ultra-relativistic motion where the boost factor $\gamma$ is large, 
\beq 
	\gamma \equiv 1/\sqrt{ - g_{\mu\nu} v^\mu v^\nu} \gg 1.
\label{eq:boost1}
\eeq 
Here, $v^\alpha \equiv dz^\alpha/dt$, $z^\alpha$ is the small mass' worldline coordinates, $g_{\mu\nu}$ is the background metric of the black hole with mass $M$,\footnote{The background spacetime does not need to be a black hole, but it must have a curvature length scale larger than the size of the small massive object for the perturbation theory to be well-defined.} and $t$ is the coordinate time of an observer's frame.\footnote{We remark that a natural coordinate time in a black hole spacetime is defined with respect to the asymptotically flat region where observers reside with gravitational wave detectors.
It is important to recall that the frame-dependence of the boost factor does not preclude one from studying ultra-relativistic motion \emph{relative to a given frame}. We comment on the case of massless particles later on.} As we mentioned, one of several ways to achieve a large boost factor is to imagine the mass $m$ on a bound orbit near the light ring in Schwarzschild spacetime, but our analysis is not limited to this particular scenario. 

We next find the scaling rules of various leading order quantities. The orbital frequency is related to the wavelength of the gravitational radiation through \beq \omega_{\rm orb} = d\phi/dt \sim 1/\lambda_{\rm gw}.\eeq Since, in the background of the large black hole, $\lambda_{\rm gw} \sim M$, it follows that $dt \sim M$ (also $dx^i \sim M$). Hence, the proper time along the object's worldline scales like \beq d\tau \sim dt/\gamma \sim M/\gamma\eeq and its four-velocity is \beq u^\alpha \equiv dz^\alpha/d\tau = \gamma v^\alpha \sim \gamma,\eeq for an ultra-relativistic motion. For the scaling of the metric perturbations $h_{\mu\nu}$ produced by this ultra-relativistic small mass $m$ we use the leading order solution
\beq
	h_{\mu\nu}(x) \sim \int_{x'}  G_{\mu\nu\alpha'\beta'}(x, x') T^{\alpha' \beta' }(x'), 
\eeq
with \beq T^{\alpha \beta} (x) \propto m\int d\tau \, \frac{ \delta^4 (x^\mu-z^\mu(\tau)) }{  \sqrt{-g} } u^{\alpha} u^{\beta},\eeq and $\int_x \equiv \int d^4x \sqrt{-g}$. We find \beq h_{\mu\nu} \sim E_m/M = \epsilon\eeq where we used $\nabla_\alpha \sim \partial_\alpha \sim 1/M$ and $G_{\mu\nu\alpha' \beta'} \sim 1/M^2$ for the scaling of the Green function in a curved background (this follows almost entirely from dimensional analysis). 
Finally, the leading order effective action scales like
\beq
S^0_{\rm pp} [z^\mu] = -m \int d\tau \sim m M/\gamma \sim L/N,
\eeq
as anticipated. The scaling rules are summarized~below:
\vskip -0.4cm
\begin{table}[h!]
\begin{eqnarray}
\nonumber
\begin{array}{|c|c|c|c|c|c|c|}
  \hline 
dz^\alpha &\nabla_\alpha & d\tau & u^\alpha & h_{\mu\nu} & G_{\mu\nu\alpha'\beta'} & S^0_{\rm pp}\\
\hline
   M & 1/M & M/\gamma & \gamma &  \epsilon =E_m /M & 1/M^2 & L/N \\
\hline
\end{array}
\end{eqnarray}
\end{table}
\vskip -0.4cm

Because of these rules, the condition that perturbation theory is under control in the ultra-relativistic limit demands not only $\epsilon$ to be small, but also \beq \lambda \equiv \epsilon \gamma^2 = \epsilon N \ll 1.\eeq The reason is simple. After including the perturbation, the point particle action is
\begin{align}
	S_{\rm pp} [z^\mu, h_{\mu\nu} ] =  - m \int d\tau \, \sqrt{ 1 -  \gamma^2 h_{\mu\nu}v^\mu v^\nu },  
\label{eq:Spp0001}
\end{align}
where we used (\ref{eq:boost1}).
According to the scaling rules, $h_{\mu\nu} \sim \epsilon$ and $v^\mu \sim 1$, we must require $\lambda \ll 1$ for the perturbation $\gamma^2 h_{\mu\nu} v^\mu v^\nu$ to be considered small with respect to the background. This is the regime of validity of our approximations. In other words, we formally take the limit \beq
{\rm Large~}N{\rm~limit:}~~ \epsilon \to 0,~N \to \infty,~{\rm with}~\lambda = \epsilon \times N~{\rm fixed~and~small}. \nn \eeq 
This is in some sense analogous to the limit taken for an infinitely boosted Schwarzschild black hole, $\gamma \to \infty$ and $m \to 0$ with $E_m = \gamma m$ fixed and small, which yields the Aichelberg-Sexl metric \cite{Aichelburg:1970dh}. Our ultra-relativistic limit requires yet another step since for a non-trivial spacetime background. 

To obtain the different scalings for the possible terms that contribute to the self-force we first need to isolate the building blocks of our Feynman diagrams and power-count each one of them. We have either worldline or bulk vertices, which we summarize next.
Using our power counting rules we have for the vertex describing the interaction of the small object $m$ with $n$ gravitational perturbations:
\begin{align}
	 \WorldlineVertex ~ \sim {} & (m) \left( \frac{ M }{ \gamma } \right) \left( \gamma^2 \right)^n = m M \gamma^{2n-1} ,
\end{align}
which arises from expanding the point particle action in (\ref{eq:Spp0001})
\beq
S_{\rm pp} =  -m \int d\tau + m \sum_{n=1}^\infty \frac{ (2n-3)!! }{ 2^n n!} \int d\tau \,  \left(h_{\alpha\beta} u^\alpha u^\beta \right)^n .
\eeq
Notice that we {\it truncate} the external legs and we do not yet include the scaling for $h_{\alpha\beta}$, which ought to be contracted with worldline or bulk couplings and will introduce an extra factor of $G_{\mu\nu\alpha'\beta'} \sim M^{-2}$ for each propagator in a given diagram. Next, we need the bulk vertices that follow from expanding the (gauge-fixed) Einstein-Hilbert action in powers of $h_{\mu\nu}$ about the given background spacetime $g_{\mu\nu}$. At $n^{\rm th}$ order this is given schematically by
\begin{align}
	S_{\rm EH} = \sum_{n=2}^\infty \int_x  \, \nabla h \nabla h \, h^{n-2},
\end{align}
where $\nabla$ indicates covariant derivatives. It is easy to show that the vertex for $n$ interacting gravitational perturbations scales as
\begin{align}
	\BulkVertex \sim M^2
 \end{align} 
in four spacetime dimensions. This completes the power-counting rules for the building blocks of the EFT formalism. To compute the {\it classical} effective action we simply need to add up all possible {\it tree-level} diagrams. (By this we mean we do not include closed gravitational loops that represent quantum effects.) The effective action then takes the form
\begin{align}
	S_{\rm eff} [ z^\mu ] = {} & \EffActionLO + \EffActionNLO + \EffActionNNLOone + \EffActionNNLOtwo  + \EffActionNNNLOone + \EffActionNNNLOtwo \nonumber \\
		& + \EffActionNNNLOthree + \EffActionNNNLOfour + \EffActionNNNLOfive + \cdots
\label{eq:seff101}
\end{align}
\indent Using the rules previously derived we can power-count each diagram in the effective action, hence their contribution to the self-force. We show next that only diagrams without bulk nonlinear interactions survive in the large $N$ limit. For that purpose it is illustrative to compare the scaling of the following diagrams, which enter to $\calo(\lambda^3)$:
\begin{align}
	& \EffActionLO \sim \frac{ L }{ N } 
	, ~~~
	 \EffActionNLO \sim \frac{ \lambda L }{ N }  
\label{eq:diagramLONLO}  
\\
	& \EffActionNNLOone \sim \frac{ \lambda^2 L }{ N } 
	, ~~~
	\EffActionNNLOtwo \sim \frac{ \lambda^2 L }{ N^2 } 
\label{eq:diagramNNLO}
\\
	& \EffActionNNNLOone \sim \EffActionNNNLOtwo \sim \frac{ \lambda^3 L}{ N }
\label{eq:diagramNNNLO1}
\\
	& \EffActionNNNLOthree \sim \frac{ \lambda^3 L }{ N^2 }, ~~~
	\EffActionNNNLOfour \sim \EffActionNNNLOfive \sim \frac{ \lambda^3 L}{ N^3 } . 
\label{eq:diagramNNNLO3}
\end{align}
We already start to see the pattern: bulk nonlinearites are suppressed in the large $N$ limit.
For a generic contribution let us consider a diagram with $N_m$ mass insertions, $N^k_v$ bulk vertices with $k$-legs, and $N_p$ propagators (including internal ones). From our power counting rules we obtain the scaling 
\beq
\label{scalep1}
 (M m/\gamma)^{N_m}~M^{2(N^{\rm tot}_v-N_p)} \gamma^{2\left(2N_p - \sum_k k V_k\right)},
\eeq 
where $N^{\rm tot}_v = \sum_k N_v^k$ is the total number of bulk vertices. Let us first look at diagrams with $N^k_v=0$. Using: 
\beq N_m+ N^{\rm tot}_v - N_p - 1 = 0,\eeq which follows from the topology of the diagrams that contribute in the classical limit, the expression in \eqref{scalep1} turns into
\beq
\label{scalep2}
\frac{L}{\gamma^2} \epsilon^{(N_m-1)}\gamma^{2(N_m-1)} = \frac{L}{N} \lambda^{(N_m-1)}~~~~(N_v^k=0),
\eeq 
and is thus a $1/N$ contribution.
From a given order in $\lambda$ (namely, $N_m$ fixed) adding bulk vertices (and internal propagators) will only introduce powers of $1/N$ (see \eqref{scalep1}) since we need at least two bulk vertices to increase the number of internal propagators and each bulk vertex has at least three legs. Intuitively this is because, for a fixed number of mass insertions, we lose powers of $N$ from propagators attached to {\it two} worldline couplings, which are promoted to a bulk interaction. This is transparent in the terms depicted in \eqref{eq:diagramLONLO}-\eqref{eq:diagramNNNLO3}.

\section{Gravitational self-force in the large $N$ limit}
\label{sec:largeN}

Self-force effects in EMRIs are intrinsically non-local, depending on the past history of the small object's motion around the larger black hole. 
Capturing these real-time dissipative interactions with an (effective) action requires a careful handling of Hamilton's variational principle of stationary action so that it is consistent with initial value data for open system dynamics (i.e., the motion of the small mass). This issue was emphasized in \cite{Galley:2009px} where it was motivated by the classical limit of the ``in-in'' formalism \cite{Schwinger:1960qe}. 
A rigorous framework to handle this in a completely general (classical) context was developed in \cite{Galley:2012hx} and applied to derive radiation reaction forces through 3.5 post-Newtonian order using the EFT method in \cite{Galley:2012qs} and to viscous hydrodynamics in \cite{Grozdanov:2013dba}.\footnote{See also \cite{Endlich:2012vt} for an alternative approach.} We elaborate on the details of this construction for the self-force problem in the large $N$ limit in Appendix \ref{app:nonconservative}.

As we have shown, in the ultra-relativistic limit we can ignore all self-interactions of the metric perturbation that do not happen on the worldline. This means that the action for the small mass object and the metric perturbations can be taken as
\begin{align}
	S [ z^\mu, h_{\mu\nu} ] = {} & - \frac{ 1}{64\pi} \int_x \bigg( h_{\alpha \beta ; \mu} h^{\alpha \beta; \mu} - \frac{1}{2} h_{;\mu} h^{;\mu} \bigg)   - m \int d\tau \, \sqrt{ 1 - h_{\alpha \beta} u^\alpha u^\beta},
\label{eq:action1}
\end{align}
where we fix the Lorenz gauge for trace-reversed perturbations. For the reader worrying about finite size effects, for example (neglecting spin) terms like \cite{Galley:2008ih}
\beq C_E \int d\tau \, {\cal E}_{\alpha \beta} {\cal E}^{\alpha \beta},\eeq 
one can easily show are highly suppressed in the large $N$ expansion, first entering at $\calo(\lambda^4 L / N^5)$. This has important consequences in the regularization of the theory because, as we shall argue, we will not encounter logarithmic divergences but only power-law, which will be handled via dimensional regularization (and set to zero since they involve scaleless integrals). We briefly discuss below the general procedure for calculating the relevant diagrams in the ultra-relativistic limit, which closely follows the analysis for a nonlinear scalar field model of EMRIs in \cite{Galley:2011te}. The details of the calculation are given in Appendices \ref{app:nonconservative}--\ref{app:feyndiagrams}.

Computing the surviving diagrams in the effective action, or the diagrams for the one-point function $h_{\mu\nu}(x)$ below, involves worldline integrals over the retarded propagator, 
\begin{align}
	I (z^{\mu'}) \equiv u^{\alpha'} \!\! u^{\beta'} \!\! \int_{-\infty}^\infty \!\!\! d\tau'' \, G^{\rm ret} _{\alpha' \beta' \gamma'' \delta'' } (z^{\mu'}, z^{\mu''}) u^{\gamma''} \!\! u^{\delta''},
\label{eq:singint1}
\end{align}
which are in general divergent. Here, a prime on an index indicates the point or proper time that the quantity is being evaluated, e.g. $u^{\alpha'} = u^\alpha (\tau')$, $u^{\gamma''} = u^\gamma (\tau'')$, etc. Following \cite{Detweiler:2002mi} we split this expression into a regular $G^{R} _{\alpha' \beta' \gamma'' \delta'' }$ and singular $G^{S} _{\alpha' \beta' \gamma'' \delta'' }$ piece, which allows us to isolate the part of (\ref{eq:singint1}) that produces the divergences. It is useful to write the singular integrals in a momentum space representation, which can be given whenever the two points on the worldline can be connected by a unique geodesic\footnote{This follows because normal coordinates are usually used to coordinatize the normal neighborhood of a point $x$. The momenta vectors are conjugate to the normal coordinates and are defined only in the tangent space at $x$ so that the momentum space representation in a curved spacetime is valid only within a normal neighborhood \cite{Bunch:1979uk, Galley:PhD, Galley:momentum}.}. Using the above decomposition one writes (\ref{eq:singint1}) as \beq I(z^{\mu'}):=I_S (z^{\mu'}) + I_R (z^{\mu'})\eeq 
where the singular and regular parts are, respectively, given by
\begin{align}
	I_S (z^{\mu'}) \equiv {} & 4 u^{\alpha'} \!\! u^{\beta'} \! P_{\alpha' \beta' \gamma' \delta'} (z^{\mu'}) {\rm Re} \int_{-\infty}^\infty d\tau'' \, u_{||}^{\gamma'} u_{||}^{\delta'}  \int_{-\infty}^\infty \frac{ d^dk}{ (2\pi)^d } \, \frac{ e^{- i k^0 (\tau''- \tau') } }{ (k^0)^2 - \vec{k}^2 + i \epsilon }
\label{eq:singint2}
\end{align}
and
\begin{align}
	I_R (z^{\mu'})  \equiv I(z^\mu ) - I_S(z^\mu) = {} & u^{\alpha'} \!\! u^{\beta'} \!\! \int d\tau'' \, D^R_{\alpha' \beta' \gamma'' \delta''} (z^{\mu'} , z^{\mu''}) u^{\gamma''} \!\! u^{\delta''}
\label{eq:regint1}
\end{align}
where
\begin{align}
D^R_{\alpha' \beta' \gamma'' \delta''} (z^{\mu'} , z^{\mu''}) = {} & \Theta(\tau'' - \tau_{\rm out}) \Theta(\tau_{\rm in} - \tau'') G^{\rm ret} _{\alpha' \beta' \gamma'' \delta''} (z^{\mu'} , z^{\mu''}) \nonumber \\
	& + \Theta(\tau_{\rm out} - \tau'') \Theta( \tau'' - \tau_{\rm in})  G^R _{\alpha' \beta' \gamma'' \delta''} (z^{\mu'} , z^{\mu''})   .
\label{eq:DR1}
\end{align}
See Appendix \ref{app:green} for further details.
The singular integral in (\ref{eq:singint2}) is written in $d$ spacetime dimensions in momentum space where the momenta are dual to Fermi normal coordinates, and $u_{||}^{\gamma'} \equiv g^{\gamma'}{}_{\lambda''} (z^{\mu'}, z^{\mu''}) u^{\lambda''}$ is the result of parallel propagating the velocity vector $u^{\lambda''}$ at $z^\mu (\tau'')$ to $z^\mu(\tau')$ using the propagator of parallel transport $g^{\gamma'}{}_{\lambda''} (z^{\mu'}, z^{\mu''})$. Also, $\tau_{\rm in}$ (and $\tau_{\rm out}$) are the proper time values at which the worldline enters (and leaves) the normal neighborhood of $z^\mu (\tau')$. See \cite{Poisson:2011nh} for more details about bi-tensor calculus and Figure \ref{fig:nbhd} in Appendix \ref{app:feyndiagrams} for a cartoon picture of the normal neighborhood. As we mentioned, the singular term in (\ref{eq:singint2}) is easily shown to vanish in dimensional regularization because it is a (scale-independent) power-law divergent integral. As a consequence, the regularization of the theory becomes straightforward in the large $N$ limit. (See Appendix \ref{app:dimreg} for a proof that using dimensional regularization for evaluating the worldline integrals amounts to replacing $G^{\rm ret}_{\alpha \beta \gamma' \delta'}$ by $D^R_{\alpha \beta \gamma' \delta'}$ at any order in perturbation theory.)\\

As outlined in \cite{Galley:2011te}, in a theory that has only worldline interactions as the relevant couplings, it is simpler to compute the metric perturbations at a field point, $h_{\mu\nu}(x)$, rather than the effective action since we can simply substitute the resulting regular part of the perturbative expression into the worldline equations of motion (renormalizing parameters if necessary) to compute the self-force. In addition, computing the metric perturbations radiated by the system yields the physically observable gravitational waveform\footnote{To do this one evaluates the metric perturbation at future null infinity and changes to the transverse-traceless gauge \cite{Maggiore}.} detectable with gravitational wave detectors (whether ground-based or spaced-based depends on the total mass of the binary and its mass ratio). We show the results next and give details of the Feynman diagram calculations in Appendix \ref{app:feyndiagrams}.

\section{Gravitational perturbations and self-force to $\calo(\lambda^4/N)$}
\label{sec:fourthorder}

In the ultra-relativistic limit, the diagrams contributing to the one-point function are
\begin{align}
h_{\mu\nu} (x)
 = {} & \FieldLO + \FieldNLO + \FieldNNLOspider + \FieldNNLOrainbows   + \FieldNNNLOspiderthree + \FieldNNNLOspiderfour  \nonumber \\
		&  + \FieldNNNLOrainbowstwo + \FieldNNNLOrainbowsthree +  \cdots 
\label{eq:hmunu100}
\end{align} 
In the ultra-relativistic limit one can write the one-point function as the convolution with a worldline coupling
\emph{master source},
\begin{align} 
	h_{\mu\nu} (x) = \int d\tau' \, G^{\rm ret}_{\mu \nu \alpha' \beta'} (x, z^{\mu'}) \cals_R^{\alpha' \beta'} (z^{\mu'}) .
\label{eq:fullh1}
\end{align}
The master source ${\cal S}_R^{\alpha \beta}$ is completely finite and given through next-to-next-to-next-to-leading order by
\begin{align}
{\cal S}_R^{\alpha' \beta'} \!\! (z^{\mu'})  = {} &   \frac{ m }{ 2} u^{\alpha'} \!\! u^{\beta'} \bigg\{ 1 + \frac{m}{4} I_R(z^{\mu'})  + \frac{3m^2}{ 32} I^2_R(z^{\mu'})+ \frac{m^2}{16} u^{\gamma'} \!\! u^{\delta'} \!\!\! \int \! d\tau'' D^R_{\gamma' \delta' \epsilon'' \eta''} u^{\epsilon''} \!\! u^{\eta''} \! I_R(z^{\mu''}) \nonumber \\
		& {\hskip0.5in}  + \frac{3m^3 }{ 128} u^{\gamma'} \!\! u^{\delta'} \!\!\! \int \! d\tau'' D^R_{\gamma' \delta' \epsilon'' \eta''} u^{\epsilon''} \!\! u^{\eta''} \! I^2_R(z^{\mu''})+ \frac{5m^3}{128} I^3_R(z^{\mu'}) \nonumber \\
		& {\hskip0.5in} + \frac{ 3 m^3}{ 64}  I_R(z^{\mu'}) u^{\gamma'} \!\! u^{\delta'} \!\!\!\! \int \!\! d\tau'' \! D^R_{\gamma' \delta' \epsilon'' \eta''} u^{\epsilon''} \!\! u^{\eta''} \! I_R(z^{\mu''}) \nonumber \\
		& {\hskip0.5in} + \frac{ m^3}{ 64} u^{\gamma'} \!\! u^{\delta'} \!\!\! \int \! d\tau'' D^R_{\gamma' \delta' \epsilon'' \eta''} u^{\epsilon''} \!\! u^{\eta''} \!\! u^{\rho''} \!\! u^{\lambda''}  \int d\tau''' D^R_{\rho'' \lambda'' \tau''' \sigma'''} u^{\tau'''} \!\! u^{\sigma'''} \! I_R(z^{\mu'''})   \nonumber \\
		& {\hskip0.5in} + \calo( \lambda^4 ) \bigg\} + \cdots. 
\label{eq:master1}
\end{align}
The relevant diagrams are all computed in Appendix \ref{app:feyndiagrams}.\\

From the master source we may compute the regular part of the metric perturbation evaluated on the worldline, i.e. $h^R_{\mu\nu} (z^\mu)$, simply by convolving (\ref{eq:master1}) with $D^R_{\mu \nu \alpha' \beta'}$ in (\ref{eq:DR1}) to give
\begin{align}
	h^R_{\mu\nu} (z^\mu) = \int d\tau' \, D^R _{\mu\nu \alpha' \beta'} (z^\mu, z^{\mu'}) {\cal S}_R^{\alpha' \beta' } (z^{\mu'})  .
\label{eq:hR1}
\end{align}

We can now compute the self-force equations of motion in two ways: 1) through the effective action by directly computing the surviving diagrams in (\ref{eq:seff101}); or 2) by making the replacement $h_{\alpha \beta}(z^\mu) \to h^R_{\alpha\beta}(z^\mu)$ in the point particle action and deriving the equations of motion through the variation of that action and substituting in for the regular part of the field at the end. 
These two approaches were performed in a nonlinear scalar model of EMRIs in \cite{Galley:2010xn} and \cite{Galley:2011te}, respectively, and shown to be equivalent, though the latter was simpler to use. We derive the equations of motion, valid to all orders of the perturbation theory in the ultra-relativistic limit, in Appendix \ref{app:eom}. We find
\begin{align}
	& \left[ g_{\mu\nu} (1-H_R) + P_\mu{}^\lambda \big( h^R_{\lambda \nu} (1-H_R) + h^R_{\lambda \alpha} u^\alpha u^\beta h^R_{\beta \nu} \big) \right] a^\nu \nonumber \\
	 & {\hskip0.5in} = - \frac{1}{2} P_{\mu}{}^\lambda \bigg[ \big( 2 h^R_{\lambda \alpha; \beta} - h^R_{\alpha \beta; \lambda}\big) \big(1-H_R \big) + h^R_{\lambda \gamma} u^\gamma h^R_{\alpha \beta; \delta} u^\delta \bigg] u^\alpha u^\beta, 
\label{eq:sf1}
\end{align} 
where $P^{\mu\nu} \equiv g^{\mu\nu} + u^\mu u^\nu$ is a projection onto directions orthogonal to $u^\mu$, $h^R_{\mu\nu}$ is evaluated on the worldline using the master source in \eqref{eq:master1}, we have defined $H_R \equiv h^R_{\alpha \beta} u^\alpha u^\beta$, 
and absorbed a divergent piece into the mass $m$. (These divergences are set to zero in dimensional regularization. Recall that there are no other counter terms at leading order in $1/N$.) 
The formal perturbative expression for the self-force can be easily found by expanding out (\ref{eq:sf1}) to the desired order and using (\ref{eq:master1}) and (\ref{eq:hR1}). 
Combining the gravitational radiation given by (\ref{eq:fullh1}) and (\ref{eq:master1}) with the solution to (\ref{eq:sf1}) provide a complete (self-consistent) expression for the self-force in the ultra-relativistic limit through NNNLO.

\section{Conservative self-force effects for circular orbits near light ring}
\label{sec:conservative}

Consider the example of a circular orbit near the light ring of a Schwarzschild background. Let us also take time-symmetric boundary conditions for the gravitational radiation so that instead of the retarded Green function the propagator becomes 
\begin{align}
	\frac{ G^{\rm ret}_{\alpha \beta \gamma' \delta'} (x,x') + G^{\rm adv} _{\alpha \beta \gamma'\delta' }(x,x') }{ 2 } . 
\label{eq:halfrethalfadv1}
\end{align}
This approach is also taken in other works (see for instance \cite{DiazRivera:2004ik}). The singular structure of the integrals is the same as using outgoing boundary conditions for the radiation. We may thus replace the retarded Green function in all expressions that appear above by (\ref{eq:halfrethalfadv1}).
The symmetry of the system simplifies the formal expression because the regular integral in (\ref{eq:regint1}) is a function only of the orbital radius $r_o$, and moreover is independent of time as long as we consider the {\it conservative} part of the self-force. That is, for this example $I_R(z^{\mu}) = I_R(z_o^\mu)$, where $z_o^\mu$ are the worldline coordinates for a circular orbit with radius $r_o$. The master source in (\ref{eq:master1}) then simplifies drastically to $\calo (\lambda^4/N)$,
\begin{align}
	{\cal S}_{R}^{\alpha' \beta'} (z_o^\mu) = {} & \frac{ m }{2 } u_o^{\alpha'} \!\! u_o^{\beta'} \bigg[  1  + \frac{m}{4} I_R(z_o^\mu) + \frac{5m^2}{32} I_R^2(z_o^\mu)   + \frac{ m^3}{8} I_R^3(z_o^\mu) + \calo( \lambda^4 ) \bigg] + \cdots 
\label{eq:SRcirc1}
\end{align} 
which is a constant for a given circular orbital radius $r_o$.
Using the above master source we find the corresponding regularized metric perturbation $h^R_{\mu\nu}$ evaluated on the worldline to be 
\begin{align}
	h_{\mu \nu}^R (z_o^\mu) = {} &  \bigg[ 1 + \frac{ m }{ 4 } I_R(z_o^\mu) + \frac{ 5 m^2 }{ 32 } I_R^2 (z_o^\mu) + \frac{ m^3 }{ 8 } I_R^3 (z_o^\mu) + \calo(\lambda^4) \bigg]  \nonumber \\
	& \times \frac{ m }{ 2 }  \int d\tau' \, D^R_{\mu\nu \alpha' \beta' } (z_o^\mu, z_o^{\mu'}) u_o^{\alpha'} u_o^{\beta'} + \cdots
\label{eq:hRpert1001}
\end{align}

Knowing the regular part of the field allows us to derive many things. One of these is the conservative part of the self-force, which allows us to compute for instance  (the conserved quantity) $E = - t^\alpha u_\alpha$ defined by contracting the time-like Killing vector $t^\alpha$ with the full four-velocity.\footnote{Note that, even though $E$ is a conserved quantity it is not gauge invariant. Hence, it should not be confused with the binding energy of the orbit.  One can also, in principle, compute the binding energy of the orbit in the ultra-relativistic limit using the results in \cite{LeTiec:2011bk,LeTiec:2011dp,Barausse:2011dq}. We thank Alexandre Le Tiec for clarifying this to us.}  
The equation for $E$ is easily shown to be
\begin{align}
	& 1 - \frac{  r_o (r_o - 3M)  }{ (r_o - 2M)^2  } E^2  = \frac{ r_o}{2}  \frac{ ( 1 - H_R )  u^\alpha u^\beta  \nabla_r h^R_{\alpha \beta}   }{ ( 1 - H_R ) (1 +  f(r_o) h^R_{rr} )   + f(r_o) (h^R_{r\gamma} u^\gamma)^2  },
\label{eq:energy1}
\end{align}
where we used the radial component of the (non-perturbative) equation of motion in (\ref{eq:sf1}), $f(r_o) = 1-2M/r_o$, and $H_R \equiv h^R_{\alpha \beta} u^\alpha u^\beta$. (Both the $u^\alpha$ and $h^R_{\alpha \beta}$ depend implicitly on $E$.) In principle we need to expand (\ref{eq:energy1}) perturbatively in powers of $\lambda$ about the background energy $E_0$ of a circular geodesic. 

 Unfortunately, we find that there are contributions (from $h^R_{rr} (r_o)$ and $h^R_{r\alpha} (r_o) u^\alpha$), starting already at $\calo(\lambda^2/N)$, which have not yet been obtained numerically and published in the literature. To this extent, we expect our results will encourage the community to compute these terms in the future.

\section{Concluding remarks}
\label{sec:remarks}

We have introduced the large $N$ expansion for computing the gravitational self-force in the ultra-relativistic limit and shown that, at leading order in $1/N$, it reduces to a (mostly) combinatorial problem. As an example, we derived the self-force through fourth order and gave the (non-perturbative, implicit) expressions for a conserved energy for circular orbits near the Schwarzschild light ring. 
Our results are most useful the larger $\gamma$ is, provided $\lambda = N\epsilon = \gamma^3 q$ remains fixed and small. For example, in our computations, ignoring $1/N^2$ corrections requires \beq \lambda^2/N^2 < \lambda^4/N,~{\rm or}~ 1/\gamma^4 < q,\eeq while at the same time $\gamma^3 q < 1$ for perturbation theory to stay under control. Therefore, the range of validity lies somewhere between $1/\gamma^4 < m/M < 1/\gamma^3$. This window obviously increases with the less accuracy we demand. Moreover, our results are formally exact in the large $N$ limit.

The gravitational self-force has received significant attention lately due in part to some surprising agreements with numerical results outside its range of validity (formally replacing $m/M \to m M / (m+M)^2$) \cite{LeTiec:2011bk,LeTiec:2011dp,Barausse:2011dq}. These comparisons, however, only relied on leading order self-force effects and circular orbits. Our results in this paper open the door to check and improve such computations to very high orders in the large $N$ limit. As it is often the case, these approximations may shed light on the dynamics in scenarios where $\gamma$ is not significantly large and perhaps even in cases where the mass ratio is not taken to be small. We leave this road open for future work. Our results should also be useful to further calibrate semi-analytic merger models from the ultra-relativistic regime (e.g., see \cite{Akcay:2012ea}).\\

Let us finish by commenting on a more formal aspect of the ultra-relativistic limit. As it is well known, a boosted Schwarzschild black hole turns into an Aichelburg-Sexl (AS) shockwave in the ultra-relativistic limit with $E_m$ finite \cite{Aichelburg:1970dh}. One simple way to recover this solution is computing the one-point function using Polyakov's action \cite{Polchinski:1998rq}
\begin{align} 
	S_{\rm Poly} = \int d\lambda \left( \frac{ \dot z^\alpha(\lambda) \dot z_\alpha(\lambda) }{e(\lambda)} - e(\lambda) m^2 \right) \stackrel{ m = 0 }{ \longrightarrow}  \int \frac{ d\lambda} { e(\lambda) } \, \big( g_{\mu\nu}(z)+ h_{\mu\nu}(z) \big) \dot{z}^\mu \dot{z}^\nu, 
\end{align}
which is finite in the massless limit. Note that $e(\lambda)$ has dimensions of $1/{\rm mass}$. A special feature of this point particle action is that it does not introduce worldline non-linearities, only bulk-type which are present through the Einstein-Hilbert action. However, all the non-linear terms cancel out for the AS solution \cite{gipomax}, which is linear in $G_N$ \cite{Aichelburg:1970dh}. This is not the case in a black hole background (with finite mass $M$) because the shockwave can encounter its own ``echoes'' \cite{Zenginoglu:2012xe}. In fact, the diagrams that contribute to the effective action in this case are
\begin{align}
	S_{\rm eff} [z^\mu] = \EffActionLO + \EffActionNLO  + \EffActionNNLOtwo   + \EffActionNNNLOfour + \EffActionNNNLOfive + \cdots ~~~{(\rm massless)}.
\end{align}
Notice that, from the full set of diagrams that contribute to the effective action in (\ref{eq:seff101}), the diagrams in the massless case are in some sense dual to those in the ultra-relativistic limit of a massive particle, since only wordline couplings survive for the latter. This suggests the different diagrams in (\ref{eq:seff101}) may be related as we take $m \to 0$ in the large $N$ limit. If such a duality existed then this would provide some interesting insight into the nature of the self-force on massless particles, and gravitational interactions altogether. (This duality would also help to simplify some of the computations that appear at higher orders in the canonical self-force perturbation theory in the mass ratio.) Along that vein, it would be interesting to study AS shockwave dynamics in non-trivial backgrounds as another approach to the ultra-relativistic self-force, for instance, to study the dynamics of light crossing a black hole, the merger process in binary systems \cite{pomcw}, or to understand high-energy gravitational collisions \cite{Giddings:2009gj, gipomax}. (For the case of photons, it would also be instructive to compare with the geometric-optics limit of the Einstein-Maxwell equations.) While this is not the same limit studied here, it would be interesting to understand the seemingly dual relationship between both approaches and the connections (if any) between worldline and bulk non-linearities.

\appendix

\section{The causal variational principle of stationary action}
\label{app:nonconservative}

The action in (\ref{eq:action1}) can be written as
\begin{align}
	S [ h_{\mu\nu}, z^\mu ] = {} & - \frac{ 1}{ 64\pi} \int_x \bigg( h_{\alpha\beta ; \mu} h^{\alpha \beta ; \mu} - \frac{1}{2} h_{;\mu} h^{;\mu} \bigg) - m \int d\tau \nonumber \\
	&  + \sum_{n=1}^\infty \frac{1}{n! } \int_x h_{\alpha_1 \beta_1}(x) \cdots h_{\alpha_n \beta_n} (x) T^{\alpha_1 \cdots \beta_n} (x; z)
\label{eq:action3}
\end{align}
where we have expanded the square root in powers of the gravitational perturbation $h_{\alpha \beta}$ to get the last term and where
\begin{align}
	T^{\alpha_1 \cdots \beta_n} (x; z) \equiv m \, \frac{ (2n-3)!!}{ 2^n } \int d\tau \, \frac{ \delta^4 (x^\mu - z^\mu(\tau)) }{ \sqrt{-g} } \, u^{\alpha_1} (\tau) \cdots u^{\beta_n} (\tau)
\label{eq:Tn1}
\end{align}
is the coefficient of the $n^{\rm th}$ order term in the expansion of the point particle action.

Capturing dissipative effects on the motion of the small compact object from the emission of gravitational radiation requires a formulation of Hamilton's principle of stationary action that can accommodate generally non-conservative forces and interactions. The framework for such a principle is given in \cite{Galley:2012hx}, which provides a variational principle based on the specification of initial data rather than on the boundary data in time that is given for the usual formulation of Hamilton's principle \cite{Goldstein}. The essential feature of this new Hamilton's principle is that one formally doubles the degrees of freedom in the problem. In the general case of an open system that is free to exchange energy with some other set of possibly inaccessible degrees of freedom, the doubling allows one to introduce an arbitrary function $K$ that couples the doubled variables. As discussed in \cite{Galley:2012hx}, $K$ is responsible for the non-conservative interactions and forces acting on the system of interest. Here, because we begin with a system that conserves energy in total (i.e., gravitational perturbations and the worldline motion of the small compact object) we can set $K=0$. After all variations are performed we are then free to set the two sets of variables equal and identify the resulting equality as the physical variable. This is called the ``physical limit'' in \cite{Galley:2012hx}.

Doubling the variables in the ultra-relativistic problem amounts to letting $h_{\alpha \beta} \to ( h_{1\alpha \beta}, h_{2\alpha \beta})$ and $z^\mu \to (z_1^\mu, z_2^\mu)$. The action that allows for the irreversible processes of radiation emission is then given by
\begin{align}
	S [ h_A^{\mu\nu}, z_A^\mu ] \equiv S[ h_1^{\mu\nu}, z^\mu_1 ] - S[ h_2^{\mu\nu}, z_2^\mu ]
\label{eq:ininaction1}
\end{align}
where $A = 1,2$.
Substituting in (\ref{eq:action3}) into (\ref{eq:ininaction1}) gives the new action
\begin{align}
	S [ h_A^{\mu\nu}, z_A^\mu ] = {} & - \frac{ 1}{ 64\pi} \int_x \bigg( h^A_{\alpha\beta ; \mu} h_A^{\alpha \beta ; \mu} - \frac{1}{2} h^A_{;\mu} h_A^{;\mu} \bigg)  - m \int ( d\tau_1 - d\tau_2 ) \nonumber \\
		& + \sum_{n=1}^\infty \frac{1}{n!} \int_x \, h^{A_1}_{\alpha_1 \beta_1}(x) \cdots h^{A_n}_{\alpha_n \beta_n}(x) V_{A_1 \cdots A_n}^{\alpha_1 \cdots \beta_n}(x)   .
\label{eq:ininaction2}
\end{align}
where, for capitol Roman indices (called history indices) taking values in $1,2$, we have defined the proper time increments along each history as
\begin{align}
	d\tau_A \equiv d\lambda \, \sqrt{ - g_{\alpha \beta} (z_A^\mu) u_A^\alpha u_A^\beta }
\end{align}
and introduced a ``metric'' $c_{AB} = {\rm diag}(1,-1)$ to raise and lower the history indices. The worldine interactions in the last line of (\ref{eq:ininaction2}) are defined in terms of $T_A^{\alpha_1 \cdots \beta_n}\equiv T^{\alpha_1 \cdots \beta_n}(x;z_A)$ in (\ref{eq:Tn1}) through
\begin{align}
	V^{\alpha_1 \beta_1 \cdots \alpha_n \beta_n} _{A_1 \cdots A_n} ( x) \equiv d_{A_1 \cdots A_n}{}^B T_B ^{\alpha_1 \beta_1 \cdots \alpha_n \beta_n } (x)
\label{eq:V1000}
\end{align}
where the tensor $d$ is
\begin{align}
	d_{A_1 \cdots A_{n}}{}^B \equiv \left\{ 
		\begin{array}{cl}
			1 & {\rm if~} A_1 = \cdots = A_{n} = B = 1\\
			(-1)^{n+1} & {\rm if~} A_1 = \cdots = A_{n} = B = 2 \\
			0 & {\rm otherwise}
		\end{array}
	\right.
\label{eq:dtensor3}
\end{align}

The perturbative action in (\ref{eq:ininaction2}) is a scalar also with respect to the internal group of $SO(1,1)$ transformations of the history indices. In other words, the theory is covariant in both spacetime and history indices. We thus can choose a set of doubled field and worldline variables that is convenient for self-force calculations. As discussed in \cite{Galley:2012hx}, a convenient new basis is given by the transformation to ``$\pm$'' coordinates, which are simply the average and difference of the variables. For the field, these are
\begin{align}
	h_+^{\alpha \beta} & \equiv \frac{ h_1^{\alpha \beta} + h_2^{\alpha \beta} }{ 2 }  \\
	h_-^{ \alpha \beta} & \equiv h_1^{\alpha \beta} - h_2^{\alpha \beta} 
\label{eq:transformation1}
\end{align}
which can be written in the form $h_a ^{\alpha \beta} = \Lambda_a{}^A h_A^{\alpha \beta}$ where $a = +, -$ and $A=1,2$.

Transforming the worldline interaction terms in (\ref{eq:ininaction2}) from the ``$1,2$'' basis to the ``$\pm$'' basis gives
\begin{align}
	S [ h_a^{\mu\nu}, z_a^\mu ] = {} & - \frac{ 1}{ 64\pi} \int_x \bigg( h^a_{\alpha\beta ; \mu} h_a^{\alpha \beta ; \mu} - \frac{1}{2} h^a_{;\mu} h_a^{;\mu} \bigg)  - m \int ( d\tau_1 - d\tau_2 ) \nonumber \\
		& + \sum_{n=1}^\infty \frac{1}{n!} \int_x \, h^{a_1}_{\alpha_1 \beta_1}(x) \cdots h^{a_n}_{\alpha_n \beta_n}(x) V_{a_1 \cdots a_n}^{\alpha_1 \cdots \beta_n}(x)  
\label{eq:ininaction6}
\end{align}
and where we have used the tensor transformation for $d$,
\begin{align}
	d_{a_1 \cdots a_n}{}^b = \Lambda_{a_1}{}^{A_1} \cdots \Lambda_{a_n}{}^{A_n} \Lambda^b{}_B d_{A_1 \cdots A_n}{}^B  .
\label{eq:dtensor10}
\end{align}
It is useful to collect some useful identities and expressions for different ranks of the $d$ tensor,
\begin{align}
	d_{AB} & = c_{AB} 
\label{eq:didentity1} \\
	d_a{}^b & = \Lambda_a{}^A \Lambda^b{}_B c^{BC} c_{AC} = \Lambda_a{}^A \Lambda^b{}_B \delta_A{}^B = \Lambda_a{}^A \Lambda^b{}_A = \delta_a{}^b 
\label{eq:didentity2} \\
	d_{+-}{}^+ & = 1 
\label{eq:didentity3}\\
	d_{+--}{}^+ & = 1
\label{eq:didentity4}
\end{align}
The worldline vertex in the $\pm$ basis is then given by
\begin{align}
	\WorldlineVertex = \frac{ \delta^n S }{ \delta h^{a_1}_{\alpha_1\beta_1} (x) \cdots \delta h^{a_n} _{\alpha_n \beta_n} (x) } = V_{a_1 \cdots a_n} ^{\alpha_1 \beta_1 \cdots \alpha_n \beta_n } (x)  .
\end{align}

One advantage of working in the $\pm$ coordinates is that in the physical limit, defined as the limit in which $h_2^{\alpha \beta} \to h_1^{\alpha \beta} = h^{\alpha \beta}$ and likewise for the worldlines, the ``$+$'' variables go to their physical values and the ``$-$'' variables vanish. It then becomes immediately clear which contributions in a calculation survive in the physical limit. It can be shown that the only contributions that survive the physical limit when computing \emph{forces} and \emph{equations of motion} are those in the action that are linear in the ``$-$'' variable \cite{Galley:2012hx}. All terms that are nonlinear in the ``$-$'' variables do not contribute in the physical limit. We will work in the ``$\pm$'' coordinates unless otherwise noted.

\section{Green functions in curved spacetime}
\label{app:green}

Doubling the variables also doubles the number of Green functions, or propagators, that appear in the formalism. In fact, in the $\pm$ basis the $+$ variables evolve from initial data using the retarded Green function while the $-$ variables evolve using the advanced Green function from data specified at the final time \cite{Galley:2012hx}. The retarded and advanced propagators can be put into a matrix of propagators, which is given in the ``$\pm$'' basis by
\begin{align}
	G^{ab}_{\alpha \beta \gamma' \delta'} (x, x') & = \left( \begin{array}{cc}
				G_{\alpha \beta \gamma' \delta'}^{++}(x,x') & G_{\alpha \beta \gamma' \delta'}^{+-} (x,x') \\
				G_{\alpha \beta \gamma' \delta'}^{-+} (x,x') & G_{\alpha \beta \gamma' \delta'}^{--} (x,x')
			\end{array} \right) 
	 = \left( \begin{array}{cc}
				0 & G_{\alpha \beta \gamma' \delta'}^{\rm adv} (x,x') \\
				G_{\alpha \beta \gamma' \delta'}^{\rm ret} (x,x') & 0
			\end{array} \right)  .
\label{eq:green2}
\end{align}
The retarded (and advanced) propagator satisfies the following equation 
\begin{align}
	& \Box \left( G_{\alpha \beta \gamma' \delta'} ^{\rm ret} (x,x') - \frac{1}{2} g_{\alpha \beta} g^{\mu \nu} G^{\rm ret} _{\mu \nu \gamma' \delta'} (x,x')  \right) + 2 R_\alpha{}^\mu{}_\beta{}^\nu G^{\rm ret}_{\mu\nu \gamma' \delta'} (x,x')  
	\nonumber \\
	& {\hskip2in} = - 32\pi g_{\alpha (\gamma'} g_{\delta' ) \beta}  \frac{ \delta^4 (x-x') }{ g^{1/2} }
\label{eq:green3}
\end{align}
in the Lorenz gauge. It is important to observe the primes on the spacetime indices. The quantity $g^{\alpha}{}_{ \beta'} = g^{\alpha}{}_{ \beta'} (x,x')$ is the \emph{propagator of parallel transport}. This operator parallel transports a vector at $x'$ to $x$ along the unique geodesic connecting the two points. If $v^{\alpha'} (x')$ is a vector at $x'$ then it can be parallel propagated to $x$ to give a new vector at $x$ 
\begin{align}
	v_{||}^\alpha (x) \equiv g^\alpha {}_{\beta'} (x,x') v^{\beta'} (x')
\label{eq:parpropvector1}
\end{align}
that can be compared with other vectors and tensors that might also reside in the tangent space at $x$. See Ref.~\cite{Poisson:2011nh} for more details.

The wave equation in (\ref{eq:green3}) can be simplified by noting that if 
\begin{align}
	P_{\alpha \beta \gamma \delta} \equiv \frac{1}{2} \left( g_{\alpha \gamma} g_{\beta \delta} + g_{\alpha \delta} g_{\beta \gamma} - \frac{ 2 }{ d-2} \, g_{\alpha \beta} g_{\gamma \delta} \right)  , 
\end{align}
where $d$ is the spacetime dimension here,
then (\ref{eq:green3}) simplifies to
\begin{align}
	\Box  G_{\alpha \beta \gamma' \delta'} ^{\rm ret} (x,x') + 2 R_\alpha{}^\mu{}_\beta{}^\nu G^{\rm ret}_{\mu\nu \gamma' \delta'} (x,x') = - 32 \pi P_{\alpha \beta \gamma' \delta'} (x,x') \frac{ \delta^4 (x-x') }{ g^{1/2} }  
\label{eq:green5}	
\end{align}
in a vacuum spacetime (where $R_{\mu\nu} = 0$) and where
\begin{align}
	P_{\alpha \beta \gamma' \delta'} (x,x') & \equiv P_{\alpha \beta \mu \nu} g^\mu{}_{(\gamma'} g_{\delta' )} {}^ \nu  \\
		& =  \frac{1}{2} \bigg( g_{\alpha \gamma'} (x,x') g_{\beta \delta'} (x,x') + g_{\alpha \delta'} (x,x') g_{\beta \gamma'} (x,x')  - \frac{ 2 }{ d-2} \, g_{\alpha \beta} (x) g_{\gamma' \delta'} (x') \bigg)  .
\end{align}
Now, (\ref{eq:green5}) takes a more standard looking form on the left side at the expense of extra tensor structure for the Dirac delta source term on the right side.

When $x' = x$ the retarded Green function $G^{\rm ret} _{\alpha \beta \gamma' \delta'} (x,x')$ is singular and will be important when evaluating worldline integrals to compute the gravitational perturbations and the self-force on the small compact object. If $x$ and $x'$ are sufficiently close that they are connected by a \emph{unique} geodesic then $x'$ is said to lie in the \emph{normal neighborhood} of $x$. In this case, we can write down the form of the retarded Green function using Hadamard's ansatz \cite{Hadamard}. In $d=4$ dimensions Hadamard's ansatz for the retarded Green function is (see \cite{Poisson:2011nh} for more details)
\begin{align}
	G^{\rm ret} _{\alpha \beta \gamma' \delta' } (x,x') = {} & 8 \Theta_+(x, \Sigma_{x'} ) \bigg[ P_{\alpha \beta \gamma' \delta'} (x,x') \Delta^{1/2} (x,x') \delta (\sigma (x,x'))  \nonumber \\
		& {\hskip 1in} + V_{\alpha \beta \gamma' \delta' } (x,x') \Theta (-\sigma(x,x') ) \bigg]  .
\label{eq:hadamard1}
\end{align}
The individual factors require some explanation. $\Theta_+ (x, \Sigma_{x'})$ is the Heaviside step function and equals $1$ if $x$ is to the future of the spatial hypersurface containing $x'$ and $0$ otherwise. Synge's world function is $\sigma(x,x')$ and equals to half of the proper time between $x$ and $x'$ along the unique geodesic connecting them. If $x$ is time-like separated from $x'$ then $\sigma < 0$. If $x$ is space-like separated from $x'$ then $\sigma > 0$. Finally, if $x$ is light-like separated from $x'$ then $\sigma = 0$. Hence, $\Theta (-\sigma(x,x'))$ is $1$ only for time-like separated points and zero otherwise. The smooth function $V_{\alpha \beta \gamma' \delta'} (x,x')$ solves the homogeneous wave equation and is otherwise irrelevant for our purposes here as we shall see. Lastly, $\Delta (x,x')$ is the van Vleck determinant and the factor of $8$ arises from our (non-standard) normalization coming from the $32\pi$ on the right side of (\ref{eq:green5}). The factor of $P_{\alpha \beta \gamma' \delta' }(x,x')$ on the first term is a direct consequence of the former's appearance as part of the Dirac delta source term in (\ref{eq:green5}).

The interpretation of (\ref{eq:hadamard1}) is as follows. The first term is proportional to $\Theta_+(x, \Sigma_{x'}) \delta( \sigma)$ and thus has support only on the forward lightcone emanating from $x'$. The second term is proportional to $\Theta_+(x, \Sigma_{x'}) \Theta( -\sigma)$ and has support on and \emph{within} the forward lightcone. Therefore, the second term accounts for all the backscattering of the wave as it propagates in curved spacetime while the first term describes the singular propagation on the lightcone.

The retarded Green function can be decomposed arbitrarily into regular and singular pieces. However, a convenient decomposition is to use the one introduced by Detweiler and Whiting in \cite{Detweiler:2002mi},
\begin{align}
	G^{\rm ret} _{\alpha \beta \gamma' \delta' } (x,x') = G^{R} _{\alpha \beta \gamma' \delta' } (x,x') + G^{S} _{\alpha \beta \gamma' \delta' } (x,x')  .
\label{eq:DW1}
\end{align}
The regular ($R$) and singular ($S$) pieces have the following properties. The regular part satisfies the homogeneous wave equation, is regular everywhere on the worldline, and is the part of the retarded propagator that is actually responsible for exerting the self-force on the small compact object. The singular part satisfies the inhomogeneous wave equation, carries all of the divergent structure of the retarded propagator, and exerts absolutely no force on the small compact object.
When $x'$ is in the normal neighborhood of $x$ we can write the regular and singular parts of the retarded propagator from Hadamard's ansatz in (\ref{eq:hadamard1}) as \cite{Poisson:2011nh}
\begin{align}
	G^R _{\alpha \beta \gamma' \delta' } (x,x') = {} & 4 P_{\alpha \beta \gamma' \delta'} (x,x') \Delta^{1/2} (x,x') \delta (\sigma) \big[ \Theta_+ (x, \Sigma_{x'} ) - \Theta_- ( x, \Sigma_{x'}) \big] \nonumber \\
		& + 8 V_{\alpha \beta \gamma' \delta' } (x,x') \Theta ( -\sigma)  
\label{eq:GR1}  \\
	G^S _{\alpha \beta \gamma' \delta' } (x,x') = {} & 4 P_{\alpha \beta \gamma' \delta'} (x,x') \Delta^{1/2} (x,x') \delta (\sigma) - 4 V_{\alpha \beta \gamma' \delta' } (x,x') \Theta ( \sigma)   .
\label{eq:GS1}	
\end{align}

We will be using dimensional regularization to regularize the divergent worldline integrals that will be encountered in Appendix \ref{app:feyndiagrams}. Dimensional regularization analytically continues the value of an integral in the complex space of dimensions. Since the above expressions for the Green functions are given specifically in $d=4$ then these will not be in a useful form for dimensional regularization. However, we can give a momentum space representation for the singular piece, which is all we really need in order to carry out dimensional regularization. Let us evaluate (\ref{eq:GS1}) with both points on the worldline. Because the worldline is a time-like curve it follows that $\sigma (z^\mu, z^{\mu'}) < 0$ where $z^\mu \equiv z^\mu(\tau)$ and $z^{\mu'} \equiv z^{\mu} (\tau')$. From this we have that the second term in (\ref{eq:GS1}) gives no contribution and
\begin{align}
	G^S _{\alpha \beta \gamma' \delta' } (z^\mu, z^{\mu'} ) = {} & 4 P_{\alpha \beta \gamma' \delta'} (z^\mu, z^{\mu'}) \Delta^{1/2} (z^{\mu}, z^{\mu'} ) \delta (\sigma (z^{\mu}, z^{\mu'})  .
\label{eq:GS2}
\end{align}
The delta function enforces that there is a contribution only when $\tau' = \tau$. As there are no derivatives in the worldline vertices, which are the only ones that contribute to any of the diagrams in the ultrarelativistic limit, then we do not have to worry about derivatives acting on our propagators and we can immediately set $\tau' = \tau$ in which case $\Delta^{1/2}$ equals to $1$ giving
\begin{align}
	G^S _{\alpha \beta \gamma' \delta' } (z^\mu, z^{\mu'} ) = {} & 4 P_{\alpha \beta \gamma' \delta'} (z^\mu, z^{\mu'}) \delta (\sigma (z^{\mu}, z^{\mu'})  .
\label{eq:GS3}
\end{align}
(We could set $\tau' = \tau$ in the $P$ factor but we have to be careful to keep the proper index structure so that the singular propagator transforms as a rank-2 tensor at both $x$ and $x'$.)

Let $s \equiv \tau' - \tau$.
In Fermi normal coordinates $s$ measures the proper time along the worldline (imagine that $\tau$ is fixed and defines the origin of the Fermi normal time coordinate). Therefore, Synge's world function can be written simply a $\sigma (z^\mu, z^{\mu'} ) = - s^2 / 2$
and the delta function in (\ref{eq:GS3}) is
\begin{align}
	\delta (\sigma (z^{\mu}, z^{\mu'} ) ) = \frac{ \delta ( s) }{ |s| }  .
\end{align}
Notice that the form of this delta function is exactly what appears in the real part of the Feynman propagator in \emph{flat} spacetime when $\vecx = \vecx'$. We can therefore immediately write down the momentum space representation of the singular propagator in Fermi normal coordinates,
\begin{align}
	G^S_{\alpha \beta \gamma' \delta' } (z^\mu, z^{\mu'} ) = 4 P_{\alpha \beta \gamma' \delta'} (z^\mu, z^{\mu'} ) \, {\rm Re} \int \frac{ d^4 k }{ (2\pi)^4 } \frac{ e^{- i k^0 s } }{ (k^0)^2 - \veck^2 + i \epsilon }   .
\label{eq:GS4}
\end{align}
In this form, we can immediately extend the representation in (\ref{eq:GS4}) to $d$ dimensions,
\begin{align}
	G^S_{\alpha \beta \gamma' \delta' } (z^\mu, z^{\mu'} ) = 4 P_{\alpha \beta \gamma' \delta'} (z^\mu, z^{\mu'} ) \, {\rm Re} \int \frac{ d^d k }{ (2\pi)^d } \frac{ e^{- i k^0 s } }{ (k^0)^2 - \veck^2 + i \epsilon }   .
\label{eq:GS5} 
\end{align}

The momentum space representation is, of course, only valid in the normal neighborhood where $x$ and $x'$ can be connected by a unique geodesic because the momenta are dual to normal coordinates, which are only defined in the normal neighborhood. 
A momentum space representation for the Feynman Green function in curved spacetimes, including corrections from the spacetime curvature, was given originally by Bunch and Parker in \cite{Bunch:1979uk} and extended to retarded and advanced Green functions as well as to the gravitational Green functions in \cite{Galley:PhD} and \cite{Galley:momentum}.

\section{Feynman rules}
\label{app:feynrules}

We give here the Feynman rules used to translate Feynman diagrams into contributions to the effective action (and/or metric perturbation). In the ultra-relativistic limit, these rules are straightforward to derive, and are as follows:
\begin{itemize}
	\item For each worldline vertex, write down a factor of $V^{\alpha_1 \cdots \beta_n}_{a_1 \cdots b_n} (x)$.
	\item For each wavy line (propagator) connecting two points on a worldline, write down a factor of $G^{ab}_{\alpha \beta \gamma' \delta'} (x,x')$ connecting a worldline vertex with a history label $a$ and spacetime indices $\alpha, \beta$ at $x$ to another worldline vertex with a history label $b$ and spacetime indices $\gamma', \delta'$ at $x'$.
	\item For each wavy line with one external end, write down a factor of $G^{-a}_{\mu\nu \alpha' \beta'}(x,x')$. The ``$-$'' chooses the outgoing boundary condition on the radiation field by choosing the retarded Green function. (A ``$+$'' would choose the advanced Green function.)
	\item Integrate over all coordinates $x, x', \ldots$.
	\item Divide by the appropriate symmetry factor.
\end{itemize}

As an example, consider the third diagram in (\ref{eq:hmunu100}). To translate the diagram into an actual expression we start by writing down the external propagator and then glue together the worldine vertices with propagators connecting them, giving
\begin{align}
	\FieldNNLOspider = {} & \frac{1}{2!} \int_{x', x'', x'''} G^{-a}_{\mu\nu \alpha'\beta'} (x,x') V_{abc}^{\alpha'\beta' \gamma' \delta' \epsilon' \eta'} (x') G^{bd}_{\gamma' \delta' \rho'' \lambda''} (x', x'') V_d^{\rho''\lambda''}(x'') \nonumber \\
		& {\hskip0.6in} \times G^{ce}_{\epsilon' \eta' \theta''' \phi'''} (x', x''') V_d^{\theta''' \phi'''}(x''').
\end{align}
The factor of $1/2!$ is the symmetry factor of the diagram because there are two ways to write down the propagators that couple to the worldline vertices.

\section{Feynman diagram calculations of gravitational perturbations}
\label{app:feyndiagrams}

In this Appendix we give the details for calculating the gravitational perturbation $h_{\alpha \beta}(x)$ using Feynman diagrams given in (\ref{eq:fullh1}) and reproduced here with lowercase Roman letters indicating the indices for the doubled variables in the $\pm$ basis, as discussed in Appendix \ref{app:nonconservative}, 
\begin{align}
	h_{\mu\nu} (x)
 = {} & \FieldLO + \FieldNLO + \FieldNNLOspider + \FieldNNLOrainbows   + \FieldNNNLOspiderthree + \FieldNNNLOspiderfour  \nonumber \\
		&  + \FieldNNNLOrainbowstwo + \FieldNNNLOrainbowsthree +  \calo( \lambda^5/N)  .
\label{eq:onepoint1}
\end{align}
It is implied, according to the Feynman rules given in the previous appendix, that each external line carries a ``$-$'' label on its free end (i.e., the end not connected to the worldline or another propagator line) and thus picks the correct causal boundary conditions for the gravitational perturbations. These diagrams each have one external leg that is labeled by ``$-$'' because the retarded propagator in (\ref{eq:green2}) is the $-+$ component of the propagator matrix, 
\begin{align}
	G^{-+}_{\alpha \beta \gamma' \delta'} (x,x') = G^{\rm ret} _{\alpha \beta \gamma' \delta' } (x,x')  .
\end{align}
The only other option would be for the free end of each propagator leg to be ``$+$'' but that would imply using the advanced propagator, which satisfies acausal and unphysical boundary conditions.

Below, we show how to compute each diagram in (\ref{eq:onepoint1}). The first diagram is leading order (LO) in $\lambda/N$ and is trivial to calculate, as we shall show. The second diagram is next-to-leading (NLO) order and contains a divergent worldline integral that will be discussed in great detail. The remaining diagrams will be computed with less detail because regularizing the divergent worldline integrals that appear can be handled using the results in Appendix \ref{app:dimreg} for any singular worldline integral that appears in the ultra-relativistic limit.

\subsection{Leading order}

Writing down the LO diagram, the first in  (\ref{eq:onepoint1}), is a straightforward exercise in applying the Feynman rules from Appendix \ref{app:feynrules}, which gives
\begin{align}
	\FieldLO ~ = \int d^4x' \, G^{-a}_{\mu\nu \alpha' \beta'} (x,x') V_a^{\alpha' \beta'} (x')  .
\label{eq:onepoint1a}
\end{align}
Summing over the history index $a$, noting that only $G^{-+}$ contributes, and expressing $V_+^{\alpha'\beta'}$ in terms of $T_a^{\alpha'\beta'}$ it follows that
\begin{align}
	\FieldLO ~ & = \int d^4x' \, G^{\rm ret}_{\mu\nu \alpha' \beta'} (x,x') d_+{}^a T_a^{\alpha' \beta'} (x') \\
		& = \int d^4x' \, G^{\rm ret}_{\mu\nu \alpha' \beta'} (x,x') \Big[ d_+{}^+ T_+^{\alpha' \beta'} (x') + d_+{}^- T_-^{\alpha' \beta'} (x') \Big]   .
\label{eq:onepoint1c}
\end{align}
The last term in brackets vanishes since $d_a{}^b = \delta_a{}^b$ from (\ref{eq:didentity2}).
Therefore, (\ref{eq:onepoint1c}) becomes
\begin{align}
	\FieldLO ~ & = \int d^4x' \, G^{\rm ret}_{\mu\nu \alpha' \beta'} (x,x') T_+^{\alpha' \beta'} (x')   .
\label{eq:onepoint1d}
\end{align}
Taking the physical limit, where all ``$-$'' variables are set to zero and ``$+$'' variables equal to the physical values gives
\begin{align}
	\FieldLO ~ = \int d^4x' \, G^{\rm ret}_{\mu\nu \alpha' \beta'} (x,x') T^{\alpha' \beta'} (x')  . 
\label{eq:onepoint1e}
\end{align}
Finally, substituting in (\ref{eq:Tn1}) with $n=1$ and integrating over the delta function in the resulting expression gives
\begin{align}
	\FieldLO ~ =  \frac{ m }{ 2 } \int d\tau'  \, G^{\rm ret}_{\mu\nu \alpha' \beta'} (x, z^{\mu'} ) \, u^{\alpha'} u^{\beta'}  .
\label{eq:onepoint1f}
\end{align}

\subsection{Next-to-leading order}

Applying the Feynman rules to the next-to-leading order contribution to the gravitational perturbation, given by the second diagram in (\ref{eq:onepoint1}), yields
\begin{align}
	\FieldNLO  = \int  d^4x' d^4x'' \, G^{-a}_{\mu\nu \alpha' \beta'} (x,x') V_{ab}^{\alpha' \beta' \gamma' \delta'} (x') G^{bc}_{\gamma' \delta' \epsilon'' \eta''} (x', x'') V_c^{\epsilon'' \eta''}  (x'' )  .
\label{eq:onepoint21a}
\end{align}
Summing over the history indices gives
\begin{align}
	\FieldNLO = \! \! \int d^4x' d^4x'' G^{-+}_{\mu\nu \alpha' \beta'} (x,x') V_{+b}^{\alpha' \beta' \gamma' \delta'} \! (x' ) G^{bc}_{\gamma' \delta' \epsilon'' \eta''} (x', x'') V_c^{\epsilon'' \eta''} \!  (x'' )  + O( -)
\label{eq:onepoint21b}
\end{align}
In doing the sum, we remark that the index on the last factor $V_c^{\epsilon'' \eta''}$ vanishes in the physical limit if $c = -$ because
\begin{align}
	V_-^{\epsilon'' \eta''} = V_1^{\epsilon'' \eta''} - V_2^{\epsilon'' \eta''}  ,
\end{align}
which vanishes in the physical limit and is indicated by $O(-)$ in (\ref{eq:onepoint21b}). Therefore, only $c=+$ gives a relevant contribution. Finishing the summation over the history indices results in
\begin{align}
	\FieldNLO & = \int d^4x' d^4x'' \, G^{\rm ret}_{\mu\nu \alpha' \beta'} (x,x') V_{+-}^{\alpha' \beta' \gamma' \delta'} (x') G^{\rm ret}_{\gamma' \delta' \epsilon'' \eta''} (x', x'') V_+^{\epsilon'' \eta''}  (x'' )  .
\label{eq:onepoint21c}
\end{align}
Next, we substitute in (\ref{eq:V1000}) evaluated in the ``$\pm$'' basis to find
\begin{align}
	\FieldNLO & = \! \! \int d^4x' d^4x'' G^{\rm ret}_{\mu\nu \alpha' \beta'} (x,x') d_{+-}{}^a T_{a}^{\alpha' \beta' \gamma' \delta'} \! (x' ) G^{\rm ret}_{\gamma' \delta' \epsilon'' \eta''} (x', x'')  d_+{}^b T_b^{\epsilon'' \eta''} \!  (x'' )
\label{eq:onepoint21d}
\end{align}
The only non-zero contributions come $d_+{}^+ = 1$ from (\ref{eq:didentity2}) and $d_{+-}{}^+ = 1$ from (\ref{eq:didentity3}) so that we are left with
\begin{align}
	\FieldNLO & = \int d^4x' d^4x'' \, G^{\rm ret}_{\mu\nu \alpha' \beta'} (x,x') T_{+}^{\alpha' \beta' \gamma' \delta'} (x' ) G^{\rm ret}_{\gamma' \delta' \epsilon'' \eta''} (x', x'')  T_+^{\epsilon'' \eta''}  (x'' ) .
\label{eq:onepoint21e}
\end{align}
Taking the physical limit, substituting in for the $T$ tensors from (\ref{eq:Tn1}), and integrating over the resulting delta functions gives
\begin{align}
	\FieldNLO & = \frac{ m^2 }{ 8 } \int \! d\tau' \, G^{\rm ret}_{\mu\nu \alpha' \beta'} (x,z^{\mu'} ) u^{\alpha'} u^{\beta'} \left( u^{\gamma'} u^{\delta'} \int d\tau'' \, G^{\rm ret}_{\gamma' \delta' \epsilon'' \eta''} (z^{\mu'}, z^{\mu''})  u^{\epsilon''} u^{\eta''} \right) .
\label{eq:onepoint21f}
\end{align}
Notice that the term in parentheses is a spacetime scalar.

The Green function in the $\tau''$ integral is evaluated with both points on the worldline and thus diverges when $\tau'' \to \tau'$. Therefore, the $\tau''$ integral has to be regularized to isolate the divergence. In flat space, one can just use a momentum space representation to carry out all calculations for the regular and singular contributions. In curved spacetime, the regularization procedure is somewhat more involved than in flat spacetime partly because an arbitrary spacetime does not admit a \emph{global} momentum space representation for the Green functions. At best, the propagators have a momentum space representation within the normal neighborhood where the two points of the Green function $x$ and $x'$ can be connected by a unique geodesic as discussed in Appendix \ref{app:green}. This means that we have to break up contributions to the Green function into two parts: those where the $x$ and $x'$ are in a normal neighborhood and those that are not. The divergence comes from the former part while the latter gives a completely finite contribution because $x$ and $x'$ are never equal. Once this decomposition is made we can use the momentum space representation in (\ref{eq:GS5}) for the singular part of the Green function in the normal neighborhood and thus use dimensional regularization for the singular worldline integral appearing in (\ref{eq:onepoint21f}) and elsewhere.

\begin{figure}
	\includegraphics[width=0.37\columnwidth]{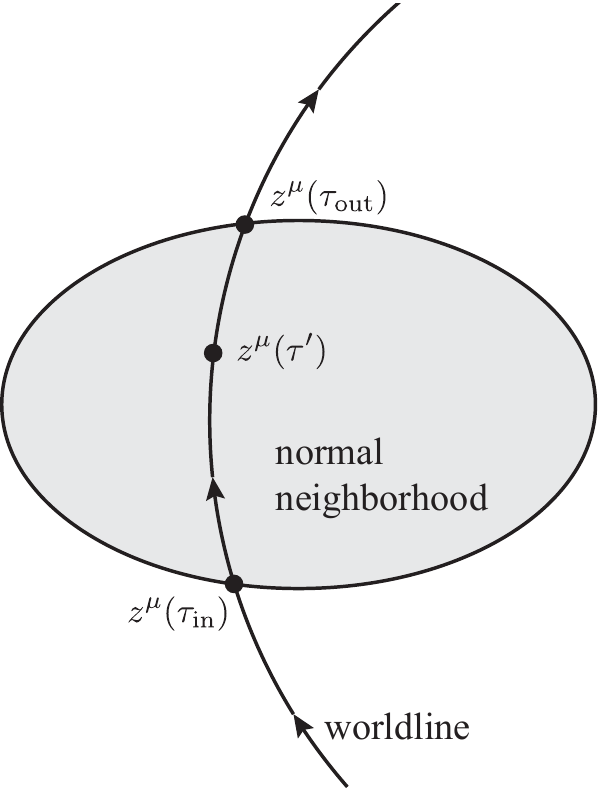}
	\caption{The normal neighborhood of $x^\mu = z^\mu(\tau')$. All points within the boundary can be connected to $x$ by a unique geodesic. Within the normal neighborhood one can construct a momentum space representation for the retarded Green function.}
	\label{fig:nbhd}
\end{figure}

The divergent part of (\ref{eq:onepoint21f}) is given by the following scalar integral
\begin{align}
	I (z^{\mu'}) \equiv u^{\gamma'} u^{\delta'} \!\! \int_{-\infty}^\infty d\tau'' \, G^{\rm ret}_{\gamma' \delta' \epsilon'' \eta''} (z^{\mu'}, z^{\mu''})  u^{\epsilon''} u^{\eta''}  .
\label{eq:integral1} 
\end{align}
Figure \ref{fig:nbhd} shows that only part of the worldline integral lies within the normal neighborhood of $z^{\mu'} = z^\mu (\tau')$. If $\tau_{\rm in}$ is the proper time that the worldline enters the normal neighborhood and $\tau_{\rm out}$ is the time it leaves then we can write (\ref{eq:integral1}) as
\begin{align}
	I(z^{\mu'}) = u^{\gamma'} u^{\delta'} \!\! \left( \int_{-\infty}^{\tau_{\rm in}} + \int_{\tau_{\rm in}} ^{\tau_{\rm out}} + \int _{\tau_{\rm out}} ^{\infty} \right) d\tau'' \, G^{\rm ret}_{\gamma' \delta' \epsilon'' \eta''} (z^{\mu'}, z^{\mu''})  u^{\epsilon''} u^{\eta''}  .
\label{eq:integral2}
\end{align}
The integral over $\tau''$ from $\tau_{\rm out}$ to $\infty$ vanishes because the retarded propagator vanishes for $\tau'' > \tau'$ leaving
\begin{align}
	I (z^{\mu'}) = u^{\gamma'} u^{\delta'} \!\! \left( \int_{-\infty}^{\tau_{\rm in}} + \int_{\tau_{\rm in}} ^{\tau_{\rm out}} \right) d\tau'' \, G^{\rm ret}_{\gamma' \delta' \epsilon'' \eta''} (z^{\mu'}, z^{\mu''})  u^{\epsilon''} u^{\eta''}  .
\label{eq:integral3}
\end{align}
Next, we write the retarded propagator in terms of Detweiler \& Whiting's singular and regular parts as in (\ref{eq:DW1}), which gives
\begin{align}
	I (z^{\mu'}) = {} & u^{\gamma'} u^{\delta'} \!\! \left( \int_{-\infty}^{\tau_{\rm in}} + \int_{\tau_{\rm in}} ^{\tau_{\rm out}} \right) d\tau'' \, G^{\rm R}_{\gamma' \delta' \epsilon'' \eta''} (z^{\mu'}, z^{\mu''})  u^{\epsilon''} u^{\eta''} \nonumber \\
		& + u^{\gamma'} u^{\delta'} \!\! \left( \int_{-\infty}^{\tau_{\rm in}} + \int_{\tau_{\rm in}} ^{\tau_{\rm out}} \right) d\tau'' \, G^{\rm S}_{\gamma' \delta' \epsilon'' \eta''} (z^{\mu'}, z^{\mu''})  u^{\epsilon''} u^{\eta''}  .
\label{eq:integral4}
\end{align}
The first line in (\ref{eq:integral4}) is completely finite. In the second line, the integral
\begin{align}
	u^{\gamma'} u^{\delta'} \!\! \int_{-\infty}^{\tau_{\rm in}} \!\! d\tau'' \, G^S_{\gamma' \delta' \epsilon'' \eta'' } (z^\mu, z^{\mu'}) u^{\epsilon''} u^{\eta''}
\end{align}
is also finite because $\tau''$ never equals $\tau'$ in this integration range. The only divergent contribution to (\ref{eq:integral4}) thus comes from the remaining integral over the singular Green function,
\begin{align}
	I_S  (z^{\mu'}) \equiv  u^{\gamma'} u^{\delta'} \!\!  \int_{\tau_{\rm in}} ^{\tau_{\rm out}} \!\!\! d\tau'' \, G^{\rm S}_{\gamma' \delta' \epsilon'' \eta''} (z^{\mu'}, z^{\mu''})  u^{\epsilon''} u^{\eta''}   .
\label{eq:IS1}
\end{align}
Because the other terms in (\ref{eq:integral4}) are all finite we can collect them together into one integral that's integrated from $-\infty$ up to $\tau'' = \tau' - \epsilon$ where $\epsilon \to 0^+$ to give the regular part of (\ref{eq:integral4}),
\begin{align}
	I_R (z^{\mu'}) \equiv {} & I(z^{\mu'}) - I_S(z^{\mu'}) \\
		= {} &  u^{\gamma'} u^{\delta'} \!\!  \int_{\tau_{\rm in}} ^{\tau_{\rm out}} \!\!\! d\tau'' \, G^{R} _{\gamma' \delta' \epsilon'' \eta''} (z^{\mu'} , z^{\mu''} ) u^{\epsilon''} u^{\eta''} +  u^{\gamma'} u^{\delta'} \!\!  \int_{-\infty} ^{\tau_{\rm in}} \!\!\! d\tau'' \, G^{\rm ret} _{\gamma' \delta' \epsilon'' \eta''} (z^{\mu'} , z^{\mu''} ) u^{\epsilon''} u^{\eta''}  .
\end{align}
Defining
\begin{align}
	D^R _{\gamma' \delta' \epsilon'' \eta''} (z^{\mu'}, z^{\mu''}) = {} & \Theta(\tau_{\rm out} - \tau'') \Theta(\tau'' - \tau_{\rm in}) G^R_{\gamma' \delta' \epsilon'' \eta''} (z^{\mu'}, z^{\mu''})  \nonumber \\
		& + \Theta(\tau_{\rm in} - \tau'') G^{\rm ret}_{\gamma' \delta' \epsilon'' \eta''} (z^{\mu'}, z^{\mu''})
\label{eq:DR1}
\end{align}
allows for $I_R(z^{\mu'})$ to be written more simply as
\begin{align}
	I_R (z^{\mu'}) \equiv {} & u^{\gamma'} u^{\delta'} \!\!  \int_{-\infty} ^{\infty} \!\!\! d\tau'' \, D^R _{\gamma' \delta' \epsilon'' \eta''} (z^{\mu'} , z^{\mu''} ) u^{\epsilon''} u^{\eta''}  .
\label{eq:regintegral10}
\end{align} 
Then (\ref{eq:integral4}) can be written as
\begin{align}
	I (z^{\mu'}) = I_R (z^{\mu'}) + I_S  (z^{\mu'})  .
\label{eq:integral5}
\end{align}

The divergence lies in the singular integral in (\ref{eq:IS1}). Since the integral is over points that lie within the normal neighborhood of $z^{\mu'}$ we can utilize the momentum space representation in (\ref{eq:GS5}) to write (\ref{eq:IS1}) as
\begin{align}
	I_S (z^{\mu'}) = {} & 4 u^{\gamma'} u^{\delta'}  {\rm Re} \int_{\tau_{\rm in}} ^{\tau_{\rm out}} \!\!\! d\tau'' \, P_{\gamma' \delta' \epsilon'' \eta''} (z^{\mu'}, z^{\mu''} ) \int \frac{ d^d k}{ (2\pi)^d } \, \frac{ e^{-i k^0 (\tau'' - \tau') } }{ (k^0)^2 - \veck^2 + i\epsilon}  .
\label{eq:IS2}
\end{align}
Recall that the singular part of the propagator is proportional to $\delta (s)$ and is thus localized in time. We may thus replace the limits of integration by $\pm \infty$ without loss of accuracy so that
\begin{align}
	I_S (z^{\mu'}) = {} & 4 u^{\gamma'} u^{\delta'} {\rm Re} \int_{-\infty} ^{\infty} d\tau'' \, P_{\gamma' \delta' \epsilon'' \eta''} (z^{\mu'}, z^{\mu''} ) u^{\epsilon''} u^{\eta''}  \int \frac{ d^d k}{ (2\pi)^d } \, \frac{ e^{-i k^0 (\tau'' - \tau' ) } }{ (k^0)^2 - \veck^2 + i\epsilon}   .
\label{eq:IS3}
\end{align}

Our next step is to expand the integrals around $\tau'' = \tau'$. It is convenient to define $s \equiv \tau'' - \tau'$ for this purpose. However, we have to take care because the expansion in a curved spacetime requires that \emph{all} tensor quantities are evaluated at the same spacetime point because then the tensors are associated with the same tangent space. The velocities $u^{\epsilon''}$ and $u^{\eta''}$ are vectors at $z^{\mu''}$ and not at the point $z^{\mu'}$ where the integral diverges. We must parallel transport these vectors to $z^{\mu'}$ before carrying out the expansion around $\tau'' = \tau'$. This is accomplished by invoking the propagator of parallel transport in (\ref{eq:parpropvector1}). Also, the tensor $P_{\gamma' \delta' \epsilon'' \eta''}$ transforms as a rank 2 symmetric tensor at $z^{\mu''}$ so that piece also needs to be parallel transported to $z^{\mu'}$. We may thus write
\begin{align}
	u^{\epsilon''} & = g^{\epsilon''}{}_{\lambda'} (z^{\mu''}, z^{\mu'}) u_{||} ^{\lambda'} \\
	u^{\eta''} & = g^{\eta''}{}_{\rho'} (z^{\mu''}, z^{\mu'}) u_{||} ^{\rho'}  , 
\end{align}
which is the inverse equation of (\ref{eq:parpropvector1}), and
\begin{align}
	P_{\gamma' \delta' \epsilon'' \eta''} (z^{\mu'}, z^{\mu''} ) = P_{\gamma' \delta' \sigma' \xi'} (z^{\mu'}) g^{\sigma'}{}_{\epsilon''} (z^{\mu'}, z^{\mu''} ) g^{\xi'}{}_{\eta''} (z^{\mu'}, z^{\mu''} )  .
\end{align}
Then, performing the contractions of $P$ with the velocities, which is needed for evaluating (\ref{eq:IS3}) gives
\begin{align}
	P_{\gamma' \delta' \epsilon'' \eta''} (z^{\mu'}, z^{\mu''} ) u^{\epsilon''} u^{\eta''} = {} & P_{\gamma' \delta' \sigma' \xi'} (z^{\mu'}) g^{\sigma'}{}_{\epsilon''} (z^{\mu'}, z^{\mu''} ) g^{\xi'}{}_{\eta''} (z^{\mu'}, z^{\mu''} ) \nonumber \\
		& \times g^{\epsilon''}{}_{\lambda'} (z^{\mu''}, z^{\mu'}) u_{||} ^{\lambda'} g^{\eta''}{}_{\rho'} (z^{\mu''}, z^{\mu'}) u_{||} ^{\rho'}  .
\end{align}
Using the identity \cite{Poisson:2011nh}
\begin{align}
	g^{\sigma'}{}_{\epsilon''} (z^{\mu'}, z^{\mu''}) g^{\epsilon''} {}_{\lambda'} (z^{\mu''}, z^{\mu'}) = g^{\sigma'}{}_{\lambda'} (z^{\mu'}) , 
\end{align}
which merely indicates the reciprocal relationship between parallel propagation from $z^{\mu''}$ to $z^{\mu'}$ and back again, it follows that
\begin{align}
	P_{\gamma' \delta' \epsilon'' \eta''} (z^{\mu'}, z^{\mu''} ) u^{\epsilon''} u^{\eta''} & = P_{\gamma' \delta' \sigma' \xi'} (z^{\mu'}) u_{||}^{\sigma'} u_{||}^{\xi'}  .
\end{align}
The singular integral in (\ref{eq:IS3}) becomes
\begin{align}
	I_S (z^{\mu'}) = {} & 4 u^{\gamma'} u^{\delta'} P_{\gamma' \delta' \sigma' \xi'} (z^{\mu'} ) {\rm Re} \int_{-\infty} ^{\infty} \!\!\! ds \, u_{||}^{\sigma'} u_{||}^{\xi'}  \int \frac{ d^d k}{ (2\pi)^d } \, \frac{ e^{-i k^0 s } }{ (k^0)^2 - \veck^2 + i\epsilon}  .
\label{eq:IS4}
\end{align}

Next, we expand $u_{||}^{\sigma'}$ about $\tau'' = \tau'$, or equivalently, about $s=0$. It can be shown that (e.g., see Appendix A in \cite{Galley:2006gs})
\begin{align}
	u_{||}^{\sigma'} & = g^{\sigma'}{}_{\epsilon''} (z^{\mu'}, z^{\mu''} ) u^{\epsilon''} \\
		& = u^{\sigma'} + s a^{\sigma'} + \frac{s^2}{2} \frac{ D a^{\sigma'} }{ d\tau'} + O(s^3)
\label{eq:uparprop10}
\end{align}
where the covariant parameter derivative $D / d\tau' \equiv u^{\alpha'} \nabla_{\alpha'}$ is evaluated at $\tau'$. The divergent integral is then seen to be generally of the form
\begin{align}
	I_S (z^{\mu'}) = {} & \sum_{n=0}^\infty c_n (z^{\mu'}) J_n
\label{eq:intform1}
\end{align}
where the $c_n(z^{\mu'})$ are scalar functions of $\tau'$ and 
\begin{align}
	J_n \equiv {\rm Re} \int_{-\infty}^\infty ds \, s^n \int \frac{d^d k }{ (2\pi) ^d } \, \frac{ e^{-i k^0 s} }{ (k^0)^2 - \veck^2 + i \epsilon }
\label{eq:J1}
\end{align}
for $n$ a non-negative integer. We can power count the integrals by assuming a cut-off momentum $\Lambda$ so that $k \sim \Lambda$ and $s \sim 1/ \Lambda$. Then the integral scales with the cut-off as $\sim \Lambda^{d-2-n}$ so that in $d=4$ this scales as $\sim \Lambda^{2-n}$. Because one power of $\Lambda^{-1}$ comes from the proper time increment $ds$ then we see that we could a power divergence for $n=0$ and a logarithmic divergence for $n=1$. For $n \ge 2$ the integral is finite and can be shown to vanish. In dimensional regularization the $n=0$ contribution vanishes because it is power divergent. 
The only thing left to consider is thus the potentially log divergent contribution from $n=1$ terms. For $n=1$ the integral (\ref{eq:J1}) is
\begin{align}
	J_1 & = {\rm Re} \int_{-\infty}^\infty ds \, s \int \frac{d^d k }{ (2\pi) ^d } \, \frac{ e^{-i k^0 s} }{ (k^0)^2 - \veck^2 + i \epsilon }
\label{eq:J2} \\
	& = {\rm Re} \int_{-\infty} ^\infty ds \, \int_{-\infty}^\infty \frac{ d^{d-1} k}{ (2\pi)^{d-1} } \int_{-\infty}^\infty \frac{ dk^0}{ 2\pi } \, i \frac{ \partial }{ \partial k^0 } \frac{ e^{-i k^0 s} }{ (k^0)^2 - \veck^2 + i \epsilon }   .
\end{align}
Integrating over $s$ gives a delta function in $k^0$ that we can integrate over to give
\begin{align}
	J_1 = {\rm Re} ~ i \! \int_{-\infty} ^\infty \frac{ d^{d-1} k }{ (2\pi)^{d-1} } \left[ \frac{ \partial }{ \partial k^0 } \frac{ 1 }{ (k^0)^2 - \veck^2 + i \epsilon } \right]_{k^0 = 0}   .
\end{align} 
The term in brackets is easily shown to vanish by taking the $k^0$ derivative and so $J_1=0$ for the potentially log divergent integral. In fact, one can see directly from the form of $J_n$ in (\ref{eq:J1}) that there is no scale associated with the integral because $J_n$ is just a (infinite) number, with no dependence on any external momenta or times. Such divergent and scaleless integrals always vanish in dimensional regularization.

Putting these pieces together it follows that the singular integral (\ref{eq:J1}) vanishes for all $n$. Therefore, the singular integral (\ref{eq:IS4}) also vanishes, $I_S(z^{\mu'}) = 0$, and (\ref{eq:integral5}) becomes
\begin{align}
	I (z^{\mu'} ) = I_R (z^{\mu'})   .
\label{eq:finiteintegral10}
\end{align}
Therefore, the regular part of the next-to-leading order contribution to the gravitational perturbation in (\ref{eq:onepoint21f}) is
\begin{align}
	\FieldNLO & = \frac{ m^2 }{ 8 } \int d\tau' \, G^{\rm ret}_{\mu\nu \alpha' \beta'} (x,z^{\mu'} ) u^{\alpha'} u^{\beta'} \bigg( u^{\gamma'} u^{\delta'} \!\! \int d\tau'' \, D^{\rm R}_{\gamma' \delta' \epsilon'' \eta''} (z^{\mu'}, z^{\mu''})  u^{\epsilon''} u^{\eta''}  \bigg)
\label{eq:onepoint21g} \\
		& = \frac{ m^2 }{ 8 } \! \int \! d\tau' \, G^{\rm ret}_{\mu\nu \alpha' \beta'} (x,z^{\mu'} ) u^{\alpha'} u^{\beta'}  I_R (z^{\mu'})  .
\label{eq:onepoint21h}
\end{align}

Singular worldline integrals appearing at any order in perturbation theory can all be shown to vanish in dimensional regularization. The proof is given in Appendix \ref{app:dimreg} and is similar to that given for a nonlinear scalar model of EMRIs (see Appendix B in \cite{Galley:2011te}). Therefore, once we write down the expressions for the diagrams using the Feynman rules we can immediately replace all retarded Green functions (that have \emph{both} points on the worldline) by their regularized part, $D^R$. (For example, from (\ref{eq:onepoint21f}) we could immediately write down (\ref{eq:onepoint21g}) or (\ref{eq:onepoint21h}).)

\subsection{Next-to-next-to-leading order}

At NNLO there are two contributions coming from the third and fourth diagrams in (\ref{eq:onepoint1}). We will consider these in turn.

Applying the Feynman rules to the third diagram in (\ref{eq:onepoint1}) gives
\begin{align}
	\FieldNNLOspider = {} & \frac{1}{2!} \int_{x', x'', x'''} G^{-a}_{\mu\nu \alpha'\beta'} (x,x') V_{abc}^{\alpha'\beta' \gamma' \delta' \epsilon' \eta'} (x') G^{bd}_{\gamma' \delta' \rho'' \lambda''} (x', x'') V_d^{\rho''\lambda''}(x'') \nonumber \\
		& {\hskip0.6in} \times G^{ce}_{\epsilon' \eta' \theta''' \phi'''} (x', x''') V_d^{\theta''' \phi'''}(x''')
\end{align}
where the $1/2!$ is a symmetry factor. We then perform the same steps as in the previous subsections for the first two diagrams in (\ref{eq:onepoint1}). We sum over the history indices in the $\pm$ basis, write the $V$ tensors in terms of the $T$ tensors, apply (\ref{eq:didentity2}) and (\ref{eq:didentity4}), take the physical limit (which is trivial), expand out the $T$ tensors using (\ref{eq:Tn1}), and finally integrate over the proper time delta functions to arrive at
\begin{align}
	\FieldNNLOspider = {} & \frac{ 3m^3 }{ 64 } \int d \tau'  \, G^{\rm ret}_{\mu\nu \alpha'\beta'} (x,z^{\mu'}) u^{\alpha'} \!\! u^{\beta'} \left( u^{\gamma'} \!\! u^{\delta'}  \int d\tau'' \, G^{\rm ret} _{\gamma' \delta' \rho'' \lambda''} (z^{\mu'}, z^{\mu''}) u^{\rho''} \!\! u^{\lambda''}  \right)^2.
\label{eq:spider10}
\end{align}
Notice that the proper time integrals factor, as might be suspected from the structure of the Feynman diagram. The quantity in parentheses is a scalar worldline integral that diverges. In fact, this integral is precisely the one in (\ref{eq:integral1}) that we encountered earlier in calculating the next-to-leading order contribution. Regularizing the integral is trivial in dimensional regularization, as already discussed in the previous subsection, so that we can immediately write down the final answer for this part of the NNLO contribution, 
\begin{align}
	\FieldNNLOspider = {} & \frac{ 3m^3 }{ 64 } \int d \tau'  \, G^{\rm ret}_{\mu\nu \alpha'\beta'} (x,z^{\mu'}) u^{\alpha'} \!\! u^{\beta'} I_R (z^{\mu'}) ^2 .
\label{eq:spider11}
\end{align}
$I_R(z^\mu)$ is given in (\ref{eq:regintegral10}).\\

The second contribution at NNLO to the gravitational perturbation $h_{\alpha \beta}(x)$ is
\begin{align}
	\FieldNNLOrainbows = {} & \int_{x', x'', x'''} G^{-a}_{\mu \nu \alpha' \beta'} (x,x') V_{ab}^{\alpha' \beta' \gamma' \delta' } (x') G^{bc}_{\gamma' \delta' \epsilon'' \eta''} (x', x'') V_{cd}^{\epsilon'' \eta'' \rho'' \sigma''} (x'') \nonumber \\
		& {\hskip0.4in}  \times G^{de}_{\rho'' \sigma'' \theta''' \phi'''} (x'', x''') V_e^{\theta''' \phi'''} (x''').
\end{align}
Performing the series of steps described in the paragraph above (\ref{eq:spider10}) gives
\begin{align}
	\FieldNNLOrainbows = {} & \frac{ m^3 }{ 32 } \int d\tau' \, G^{\rm ret}_{\mu \nu \alpha' \beta'} (x,z^{\mu'}) u^{\alpha'}  u^{\beta'} \!\! \int d\tau'' \, u^{\gamma'} u^{\delta'} G^{\rm ret}_{\gamma' \delta' \epsilon'' \eta''} (z^{\mu'}, z^{\mu''}) u^{\epsilon''} u^{\eta''} \!\! \nonumber \\
		& \times \int d\tau''' \, u^{\rho''} \!\! u^{\sigma''} G^{\rm ret}_{\rho'' \sigma'' \theta''' \phi'''} (z^{\mu''}, z^{\mu'''}) u^{\theta'''} \!\! u^{\phi'''} .
\end{align}
The integral over $\tau'''$ is just the scalar integral $I(z^{\mu''})$ encountered already in (\ref{eq:integral1}), hence
\begin{align}
	\FieldNNLOrainbows = {} & \frac{ m^3 }{ 32 } \int d\tau' \, G^{\rm ret}_{\mu \nu \alpha' \beta'} (x,z^{\mu'}) u^{\alpha'} u^{\beta'} \bigg( \int d\tau'' \, u^{\gamma'} u^{\delta'}   G^{\rm ret}_{\gamma' \delta' \epsilon'' \eta''} (z^{\mu'}, z^{\mu''}) u^{\epsilon''} u^{\eta''} I( z^{\mu''}) \bigg).
\label{eq:tworainbows1}
\end{align}
Notice that the factor in large parentheses involves the convolution of $I(z^{\mu''})$ instead of a product as with the previous diagram in (\ref{eq:spider11}).
Using dimensional regularization, the factor in parentheses in (\ref{eq:tworainbows1}) is shown to have vanishing singular pieces in Appendix \ref{app:dimreg}. Therefore, we may simply replace $I(z^{\mu''})$ by $I_R (z^{\mu''})$ and $G^{\rm ret}$ by $D^R$ giving
\begin{align}
	\FieldNNLOrainbows = {} & \frac{ m^3 }{ 32 } \! \int \! d\tau' \, G^{\rm ret}_{\mu \nu \alpha' \beta'} (x,z^{\mu'}) u^{\alpha'} u^{\beta'} \! \bigg( \! \int \! d\tau'' \,  u^{\gamma'} u^{\delta'}  \! D^R_{\gamma' \delta' \epsilon'' \eta''} (z^{\mu'}, z^{\mu''}) u^{\epsilon''} u^{\eta''} I_R( z^{\mu''}) \! \bigg) .
\label{eq:nnlorainbows10}
\end{align}

\subsection{Next-to-next-to-next-to-leading order}

Following the same steps as mentioned in the previous subsections, and using the result that all divergent worldline integrals can be replaced by their regular parts (see Appendix \ref{app:dimreg}), we have the following expressions for the four diagrams appearing at NNNLO in the gravitational perturbations:
\begin{align}
	 \FieldNNNLOspiderthree = {} & \frac{ 5 m^4 }{ 256} \! \int \! d\tau' \, G^{\rm ret}_{\mu\nu \alpha' \beta'} (x, z^{\mu'}) u^{\alpha'} u^{\beta'} I_R(z^{\mu'})^3 \\
	 \FieldNNNLOspiderfour = {} & \frac{ 3 m^4 }{ 256 } \! \int \! d\tau' \, G^{\rm ret}_{\mu\nu \alpha' \beta'} (x, z^{\mu'}) u^{\alpha'} u^{\beta'} \!\!\! \int \! d\tau'' \, u^{\gamma'} u^{\delta'} D^R_{\gamma' \delta' \epsilon'' \eta''} (z^{\mu'}, z^{\mu''}) u^{\epsilon''} u^{\eta''} \nonumber \\
	 	& {\hskip0.5in} \times I_R(z^{\mu''})^2 \\
	 \FieldNNNLOrainbowstwo = {} & \frac{ 3 m^4 }{ 128 } \! \int \! d\tau' \, G^{\rm ret} _{\mu\nu \alpha' \beta'} (x, z^{\mu'}) u^{\alpha'} u^{\beta'} I_R(z^{\mu'}) \nonumber \\
	 	& {\hskip0.5in} \times  \int \! d\tau'' \, u^{\gamma'} u^{\delta'} D^R _{\gamma' \delta' \epsilon'' \eta''} (z^{\mu'}, z^{\mu''}) u^{\epsilon''} u^{\eta''} I_R (z^{\mu''})  \\
	\FieldNNNLOrainbowsthree = {} & \frac{ m^4 }{ 128} \! \int \! d\tau' \, G^{\rm ret} _{\mu\nu \alpha' \beta'} (x, z^{\mu'}) u^{\alpha'} u^{\beta'} \!\!\! \int \! d\tau'' \, u^{\gamma'} u^{\delta'} D^R_{\gamma' \delta' \epsilon'' \eta''} (z^{\mu'}, z^{\mu''}) u^{\epsilon''} u^{\eta''} \nonumber \\
		&  {\hskip0.5in} \times \int \! d\tau''' \, u^{\rho''} u^{\lambda''} D^R _{\rho'' \lambda'' \tau''' \sigma'''} (z^{\mu''}, z^{\mu'''}) u^{\tau'''} u^{\sigma'''} I_R (z^{\mu'''})
\end{align} 
One may compute higher order contributions to the gravitational perturbation $h_{\mu \nu}(x)$, which turns into a combinatorial problem.

\subsection{Gravitational perturbations to NNNLO}

Putting together all the Feynman diagrams computed before gives the gravitational perturbation generated by the ultra-relativistic motion of the small compact object. However, notice that each of the contributions involve a common factor of
\begin{align}
	\int d\tau' \, G^{\rm ret} _{\mu\nu \alpha' \beta'} (x, z^{\mu'}) \big( \cdots \big)
\end{align}
where the terms indicated by $\cdots$ are specific to the actual diagram. The gravitational perturbation is thus of the form
\begin{align}
	h_{\mu\nu} (x) = \int d\tau' \, G^{\rm ret} _{\mu\nu \alpha' \beta'} (x, z^{\mu'}) {\cal S}_R (z^{\mu'}) 
\label{eq:handSR10}
\end{align}
where ${\cal S}_R (z^{\mu'})$ is the regular part of a quantity called the \emph{master source}, which was first introduced in the context of a nonlinear scalar model of EMRIs in \cite{Galley:2011te}, and is given in (\ref{eq:master1}). The master source acts as an effective stress energy for the small compact object when higher order corrections are accounted for in the ultra-relativistic limit. 

One advantage of identifying the master source is that we can convolve ${\cal S}_R$ with any Green function we like to compute a quantity of interest. For example, convolving the retarded Green function with the master source gives the radiated gravitational perturbation as in (\ref{eq:handSR10}). In particular, convolving the master source with the regular part of the retarded Green function gives the regular part of the metric perturbation
\begin{align}
	h^R_{\mu\nu} (x) = \int d\tau' \, D^{R} _{\mu\nu \alpha' \beta'} (x, z^{\mu'}) {\cal S}_R (z^{\mu'}) 
\end{align}
which can be evaluated on the worldline for computing the self-force in Appendix \ref{app:eom}.

\section{Dimensional regularization at any order in perturbation theory}
\label{app:dimreg}

In this Appendix, we prove that a regular expression for the master source (or the gravitational perturbation) can be found through any order in perturbation theory. We will use dimensional regularization to evaluate the singular integrals, in which case we will find they can be consistently set to zero.
From the form of the action in (\ref{eq:action1}) it is clear that, by virtue of the ultra-relativistic limit, all interactions are confined to be on the worldline. Since solutions involving integrals of retarded Green functions evaluated at the same point (e.g., $G^{\rm ret}_{\mu\nu \alpha' \beta'}(z^\mu(\tau) , z^\mu(\tau))$) cannot be generated in a classical theory then every singular contribution to the master source contains divergent integrals from the following set
\begin{align}
	 & \bigg\{ \calJ_n (z^{\mu_1})  \equiv \int \! d\tau_2 \cdots \int \! d\tau_{n} \, u^{\alpha_1} u^{\beta_1} G^{\rm ret} _{\alpha_1 \beta_1 \alpha_2 \beta_2} (z^{\mu_1}, z^{\mu_2}  ) u^{\alpha_2} u^{\beta_2} \cdots \nonumber \\
	  	& {\hskip0.9in} \cdots \times u^{\alpha_{n-1}} u^{\beta_{n-1}} G^{\rm ret} _{\alpha_{n-1} \beta_{n-1} \alpha_n \beta_n } (z^{\mu_{n-1} } , z^{\mu_n}  ) u^{\alpha_n} u^{\beta_n} \bigg\} _{n=2} ^N 
\label{genintegral1}
\end{align}
where $z^{\mu_i} \equiv z^\mu (\tau_i)$ for some positive integer $i$. The case $N=2$ contains just one integral, $\calJ_2(z^{\mu_1}) = I(z^{\mu_1})$, which was encountered in evaluating the NLO diagram in (\ref{eq:onepoint21f}) while for $N=3$ the two NNLO diagrams contain a $\calJ_3$ integral and a product of two $\calJ_2$ integrals, and so on.

Consider $\calJ_n (z^{\mu_1})$, any member in the set of (\ref{genintegral1}), and first regularize the $\tau_n$ integral. In the Detweiler-Whiting (DW) scheme one writes the retarded propagator as in (\ref{eq:DW1})
so that
\begin{align}
	\calJ_n (z^{\mu_1}) = {} & \int d\tau_2 \cdots d\tau_{n-1}  \left( \prod _{k=1}^{n-2} u^{\alpha_k} u^{\beta_k} G^{\rm ret}_{\alpha_k \beta_k \alpha_{k+1} \beta_{k+1}} (z^{\mu_k}, z^{\mu_{k+1}} ) u^{\alpha_{k+1}} u^{\beta_{k+1}}  \right) \nonumber \\
	&  \times \int d\tau_n \, u^{\alpha_{n-1}} u^{\beta_{n-1}} \Big[ D^R_{\alpha_{n-1} \beta_{n-1} \alpha_n \beta_n} (z^{\mu_{n-1}}, z^{\mu_n} ) \nonumber \\
	& {\hskip1.25in} + G^S_{\alpha_{n-1} \beta_{n-1} \alpha_n \beta_n} (z^{\mu_{n-1}}, z^{\mu_n} )  \Big] u^{\alpha_n} u^{\beta_n}  .
\end{align}
From (\ref{eq:finiteintegral10}) we recall that the $\tau_n$ integral is just equal, with dimensional regularization, to $I_R(z^{\mu_{n-1}})$ thereby yielding
\begin{align}
	\calJ_n (z^{\mu_1}) = {} & \int d\tau_2 \cdots d\tau_{n-1}  \left( \prod _{k=1}^{n-2} u^{\alpha_k} u^{\beta_k} G^{\rm ret}_{\alpha_k \beta_k \alpha_{k+1} \beta_{k+1}} (z^{\mu_k}, z^{\mu_{k+1}} ) u^{\alpha_{k+1}} u^{\beta_{k+1}}  \right) I_R(z^{\mu_{n-1}}) .
\end{align}
Now consider the $\tau_{n-1}$ integral,
\begin{align}
	J_n(z^{\mu_1}) = {} & \int d\tau_2 \cdots d\tau_{n-2}  \left( \prod _{k=1}^{n-3} u^{\alpha_k} u^{\beta_k} G^{\rm ret}_{\alpha_k \beta_k \alpha_{k+1} \beta_{k+1}} (z^{\mu_k}, z^{\mu_{k+1}} ) u^{\alpha_{k+1}} u^{\beta_{k+1}} \right) \nonumber \\
	& \times \int d\tau_{n-1} \, u^{\alpha_{n-2}} u^{\beta_{n-2}} \Big[ D^R_{\alpha_{n-2} \beta_{n-2} \alpha_{n-1} \beta_{n-1}} (z^{\mu_{n-2}}, z^{\mu_{n-1}} ) \nonumber \\
	& {\hskip1.25in} + G^S_{\alpha_{n-2} \beta_{n-2} \alpha_{n-1} \beta_{n-1}} (z^{\mu_{n-2}}, z^{\mu_{n-1}} )  \Big] u^{\alpha_{n-1}} u^{\beta_{n-1}} I_R (z^{\mu_{n-1}} )  .
\label{genintegral2}
\end{align}
The $\tau_{n-1}$ integral is then equal to
\begin{align}
	& \int d\tau_{n-1} \, u^{\alpha_{n-2}} u^{\beta_{n-2}} D^R_{\alpha_{n-2} \beta_{n-2} \alpha_{n-1} \beta_{n-1}} (z^{\mu_{n-2}}, z^{\mu_{n-1}} ) u^{\alpha_{n-1}} u^{\beta_{n-1}} I_R (z^{\mu_{n-1}} )  \nonumber \\
	& + \int d\tau_{n-1} \, u^{\alpha_{n-2}} u^{\beta_{n-2}} G^S_{\alpha_{n-2} \beta_{n-2} \alpha_{n-1} \beta_{n-1}} (z^{\mu_{n-2}}, z^{\mu_{n-1}} ) u^{\alpha_{n-1}} u^{\beta_{n-1}} I_R (z^{\mu_{n-1}} ) 
\label{genintegral4}
\end{align}
The first term is regular and finite.
The second term involves a proper time integral over the singular Green function multiplying a regular function of $\tau_{n-1}$. Since the latter is regular we may expand it in a Taylor series for $s_{n-1} \equiv \tau_{n-1} - \tau_{n-2}$ near zero, which gives
\begin{align}
	I_R (z^{\mu_{n-1}}) = I_R (z^{\mu_{n-2}}) + s_{n-1} \dot{I}_R (z^{\mu_{n-2}}) + O(s_{n-1}^2)
\end{align}
where a dot represents $d/d\tau_{n-2}$. Parallel propagating the velocities along a geodesic from $\tau_{n-1}$ to $\tau_{n-2}$ and using (\ref{eq:IS4}) and the above expression we find that the second term in (\ref{genintegral4}) equals
\begin{align}
	&  4 u^{\alpha_{n-2}} u^{\beta_{n-2}} P_{\alpha_{n-2} \beta_{n-2} \gamma_{n-2} \delta_{n-2}} (z^{\mu_{n-2}}) \, {\rm Re} \int_{-\infty}^\infty \!\!\! ds_{n-1} \,  u_{||}^{\gamma_{n-2}} u_{||}^{\delta_{n-2}} \nonumber \\
	 	& {\hskip0.5in} \times \Big( I_R (z^{\mu_{n-2}}) + s_{n-1} \dot{I}_R (z^{\mu_{n-2}}) + \cdots \Big) \int \frac{ d^d k}{ (2\pi)^d } \, \frac{ e^{-i k^0 s_{n-1} } }{ \left(k^0\right)^2 - \vec{k}^{\,2} + i \epsilon } .
\label{genintegral6}
\end{align}
Expanding out the parallel transported velocities in powers of $s_{n-1}$ as in (\ref{eq:uparprop10}) we see that this expression equals a sum of terms that are each proportional to the divergent, scaleless integrals that appeared in Appendix \ref{app:feyndiagrams} in (\ref{eq:J1}), which were shown to vanish in dimensional regularization.
Therefore, (\ref{genintegral6}) itself equals to zero
and (\ref{genintegral2}) is
\begin{align}
	\calJ_n(z^{\mu_1}) = {} & \int d\tau_2 \cdots d\tau_{n-2}  \left( \prod _{k=1}^{n-3} u^{\alpha_k} u^{\beta_k} G^{\rm ret}_{\alpha_k \beta_k \alpha_{k+1} \beta_{k+1}} (z^{\mu_k}, z^{\mu_{k+1}} ) u^{\alpha_{k+1}} u^{\beta_{k+1}} \right) \nonumber \\
	& \times \int d\tau_{n-1} \, u^{\alpha_{n-2}} u^{\beta_{n-2}} D^R_{\alpha_{n-2} \beta_{n-2} \alpha_{n-1} \beta_{n-1}} (z^{\mu_{n-2}}, z^{\mu_{n-1}} )  u^{\alpha_{n-1}} u^{\beta_{n-1}} I_R (z^{\mu_{n-1}} ) 
\end{align}
The remaining proper time integrals are evaluated in like manner. In particular, by induction it follows that
\begin{align}
	J_n(z^{\mu_1}) = \int d\tau_2 \cdots d\tau_{n}  \left( \prod _{k=1}^{n-1} u^{\alpha_k} u^{\beta_k} D^{R}_{\alpha_k \beta_k \alpha_{k+1} \beta_{k+1}} (z^{\mu_k}, z^{\mu_{k+1}} ) u^{\alpha_{k+1}} u^{\beta_{k+1}} \right)    
\end{align} 
Therefore, the regular part of the master source is constructed from the set of integrals in (\ref{genintegral1}) with $G^{\rm ret}_{\cdot \cdot \cdot \cdot}$ replaced by the regular part, $D^R_{\cdot \cdot \cdot \cdot}$. Hence, one can always find the regular part of the master source ${\cal S}_R$ to any order in perturbation theory in the ultra-relativistic limit. Actually, this result is true for any diagram that involves only worldline interactions in the perturbation theory, regardless of whether the small compact object moves ultra-relativistically or not.

\section{Worldline equations of motion to all orders in perturbation theory}
\label{app:eom}

The worldline equations of motion for a point mass moving in a curved background spacetime can formally be written down to all orders in the gravitational perturbation $h_{\mu\nu}$. Of course, a point particle generates divergences that must be regularized. In the previous Appendix, we showed that if the $h_{\mu\nu}$ is sourced only by worldline interactions then these worldline divergences can be regularized simply to give a finite result. For self-forced motion we can derive the finite self-force by writing down the formal equations of motion, making the replacement $h_{\alpha \beta} (z^\mu) \to h^R_{\alpha \beta}(z^\mu)$, and expanding to the desired order in $\lambda$ as shown in \cite{Galley:2011te}.

We start from the action for a point particle in a curved background
\begin{align}
	S_{\rm pp} [ z^\mu ] = - m \int d\lambda \, \sqrt{ -g_{\alpha \beta} (z^\mu) u^\alpha u^\beta - h_{\alpha \beta} (z^\mu) u^\alpha u^\beta }.
\label{eq:Spp10101}
\end{align}
The equations of motion are found by varying the action with respect to $z^\mu (\lambda')$ in the usual way,
\begin{align}
	0 = \frac{ \delta S_{\rm pp} }{ \delta z^\mu (\lambda') }
\end{align}
which gives
\begin{align}
	0 = - \int d\lambda \frac{\displaystyle\partial_\mu ( g_{\alpha \beta} + h_{\alpha \beta}) \frac{ d z^\alpha}{ d\lambda} \frac{ d z^\beta }{ d\lambda } \delta(\lambda' - \lambda) + 2 (g_{\alpha \beta} + h_{\alpha \beta}) \frac{ dz^\beta }{ d\lambda } \frac{ d }{ d\lambda } \big( \delta(\lambda'-\lambda) g_\mu{}^\alpha \big) }{ 2 \sqrt{ - g_{\gamma \delta} u^\gamma u^\delta - h_{\gamma \delta} u^\gamma u^\delta} } .
\end{align}
Carrying out the $\lambda$ integral, fixing $\lambda'$ to be the proper time of the worldline as defined on the background spacetime $g_{\alpha \beta}$ so that
\begin{align}
	g_{\alpha \beta} u^\alpha u^\beta = -1
\end{align}
where $u^\alpha (\tau) \equiv dz^\alpha / d\tau$, and writing the partial derivatives in terms of covariant derivatives $\nabla_\alpha$ (compatible with the background metric) gives the formal non-perturbative equations of motion
\begin{align}
	\frac{ D }{ d\tau} \bigg[ \frac{ \big( g_{\mu\alpha} (z) + h_{\mu \alpha}(z) \big) u^\alpha }{ \sqrt{ 1 - H } } \bigg] = \frac{ u^\alpha u^\beta h_{\alpha \beta ; \mu} }{ 2 \sqrt{1-H } }
\label{eq:fullppeom1}
\end{align}
where
\begin{align}
	H (z^\mu) \equiv h_{\alpha \beta} (z^\mu) u^\alpha u^\beta
\end{align}
and $D/d\tau = u^\alpha \nabla_\alpha$. A semi-colon followed by a spacetime index corresponds to the covariant derivative, ${}_{;\mu} = \nabla_\mu$.

We would like to have this equation expressed in a form similar to $m a^\mu = F^\mu$ so that the right-hand side comprises the self-force $F^\mu$ on the mass. To achieve this we simply expand out the covariant parameter derivative on the left side of (\ref{eq:fullppeom1}). After some algebra we find
\begin{align}
	& \left[ g_{\mu\nu} (1-H) + P_\mu{}^\lambda \big( h_{\lambda \nu} (1-H) + h_{\lambda \alpha} u^\alpha u^\beta h_{\beta \nu} \big) \right] a^\nu \nonumber \\
	 & {\hskip0.5in} = - \frac{1}{2} P_{\mu}{}^\lambda \bigg[ \big( 2 h_{\lambda \alpha; \beta} - h_{\alpha \beta; \lambda}\big) \big(1-H \big) + h_{\lambda \gamma} u^\gamma h_{\alpha \beta; \delta} u^\delta \bigg] u^\alpha u^\beta
\label{eq:fullppeom2}
\end{align} 
where
\begin{align}
	P_{\alpha \beta} = g_{\alpha \beta} + u_\alpha u_\beta
\end{align}
is orthogonal to the particle's four-velocity. Writing (\ref{eq:fullppeom2}) so that the acceleration is isolated on the left side requires ``inverting'' the tensor in square brackets, which can be accomplished perturbatively to the desired order in the metric perturbation.

As a check, expanding out (\ref{eq:fullppeom2}) to first order in the metric perturbation (and recalling that the LO acceleration is $0$ (a geodesic), and thus $a^\mu$ is of order $h_{\mu\nu}$), gives
\begin{align}
	a_\mu =  - \frac{1}{2} P_{\mu}{}^\lambda \big( 2 h_{\lambda \alpha; \beta} - h_{\alpha \beta ; \lambda} \big) u^\alpha u^\beta + O(h^2)
\end{align}
which is the correct (formal) expression for the first order self-force equation of motion before regularizing. Expanding through second order in the metric perturbation gives
\begin{align}
	a_\mu =  - \frac{1}{2} P_{\mu\nu} \big( g^{\nu\lambda} - h^{\nu \lambda} \big) \big( 2 h_{\lambda \alpha; \beta} - h_{\alpha \beta ; \lambda} \big) u^\alpha u^\beta + O(h^3)
\end{align}
As already discussed, regularizing the point particle divergences in the worldline equations of motion amounts to replacing $h_{\mu\nu} (z)$ by it's regular part $h^R_{\mu\nu}(z)$ to the given order in $\lambda$ \cite{Galley:2011te}.

\acknowledgments

We thank Alexandre Le Tiec for helpful discussions and the organizers and participants of the workshop ``Chirps, Mergers, and Explosions:\;The Final Moments of Coalescing Binaries'' held at the KITP where this work originated (NSF Grant No.\;PHY05-25915). C.R.G.\;was supported by NSF grant PHY-1068881 and CAREER grant PHY-0956189; R.A.P.\;was supported by NSF grant AST-0807444 and DOE grant DE-FG02-90ER40542.


\bibliographystyle{JHEP}
\bibliography{eft}

\end{document}